\documentclass[reprint,aps,prd,amsmath,floatfix,superscriptaddress,showpacs]{revtex4-1}

\usepackage{graphicx}
\usepackage{dcolumn}
\usepackage{bm}

\begin{document}

\pagestyle{plain}

\begin{flushleft}
FERMILAB-PUB-12-468-E, arXiv:1208.2915 [hep-ex]
\end{flushleft}


\title{Measurements of atmospheric neutrinos and antineutrinos in the MINOS Far Detector}


\newcommand{\Berkeley}{Lawrence Berkeley National Laboratory, Berkeley, California, 94720 USA}
\newcommand{\Cambridge}{Cavendish Laboratory, University of Cambridge, Madingley Road, Cambridge CB3 0HE, United Kingdom}
\newcommand{\FNAL}{Fermi National Accelerator Laboratory, Batavia, Illinois 60510, USA}
\newcommand{\RAL}{Rutherford Appleton Laboratory, Science and Technologies Facilities Council, OX11 0QX, United Kingdom}
\newcommand{\UCL}{Department of Physics and Astronomy, University College London, Gower Street, London WC1E 6BT, United Kingdom}
\newcommand{\Caltech}{Lauritsen Laboratory, California Institute of Technology, Pasadena, California 91125, USA}
\newcommand{\Alabama}{Department of Physics and Astronomy, University of Alabama, Tuscaloosa, Alabama 35487, USA}
\newcommand{\ANL}{Argonne National Laboratory, Argonne, Illinois 60439, USA}
\newcommand{\Athens}{Department of Physics, University of Athens, GR-15771 Athens, Greece}
\newcommand{\NTUAthens}{Department of Physics, National Tech. University of Athens, GR-15780 Athens, Greece}
\newcommand{\Benedictine}{Physics Department, Benedictine University, Lisle, Illinois 60532, USA}
\newcommand{\BNL}{Brookhaven National Laboratory, Upton, New York 11973, USA}
\newcommand{\CdF}{APC -- Universit\'{e} Paris 7 Denis Diderot, 10, rue Alice Domon et L\'{e}onie Duquet, F-75205 Paris Cedex 13, France}
\newcommand{\Cleveland}{Cleveland Clinic, Cleveland, Ohio 44195, USA}
\newcommand{\Delhi}{Department of Physics \& Astrophysics, University of Delhi, Delhi 110007, India}
\newcommand{\GEHealth}{GE Healthcare, Florence South Carolina 29501, USA}
\newcommand{\Harvard}{Department of Physics, Harvard University, Cambridge, Massachusetts 02138, USA}
\newcommand{\HolyCross}{Holy Cross College, Notre Dame, Indiana 46556, USA}
\newcommand{\Houston}{Department of Physics, University of Houston, Houston, Texas 77204, USA}
\newcommand{\IIT}{Department of Physics, Illinois Institute of Technology, Chicago, Illinois 60616, USA}
\newcommand{\Iowa}{Department of Physics and Astronomy, Iowa State University, Ames, Iowa 50011 USA}
\newcommand{\Indiana}{Indiana University, Bloomington, Indiana 47405, USA}
\newcommand{\ITEP}{High Energy Experimental Physics Department, ITEP, B. Cheremushkinskaya, 25, 117218 Moscow, Russia}
\newcommand{\JMU}{Physics Department, James Madison University, Harrisonburg, Virginia 22807, USA}
\newcommand{\LASL}{Nuclear Nonproliferation Division, Threat Reduction Directorate, Los Alamos National Laboratory, Los Alamos, New Mexico 87545, USA}
\newcommand{\Lebedev}{Nuclear Physics Department, Lebedev Physical Institute, Leninsky Prospect 53, 119991 Moscow, Russia}
\newcommand{\LLL}{Lawrence Livermore National Laboratory, Livermore, California 94550, USA}
\newcommand{\LosAlamos}{Los Alamos National Laboratory, Los Alamos, New Mexico 87545, USA}
\newcommand{\MIT}{Lincoln Laboratory, Massachusetts Institute of Technology, Lexington, Massachusetts 02420, USA}
\newcommand{\Minnesota}{University of Minnesota, Minneapolis, Minnesota 55455, USA}
\newcommand{\Crookston}{Math, Science and Technology Department, University of Minnesota -- Crookston, Crookston, Minnesota 56716, USA}
\newcommand{\Duluth}{Department of Physics, University of Minnesota -- Duluth, Duluth, Minnesota 55812, USA}
\newcommand{\Ohio}{Center for Cosmology and Astro Particle Physics, Ohio State University, Columbus, Ohio 43210 USA}
\newcommand{\Otterbein}{Otterbein College, Westerville, Ohio 43081, USA}
\newcommand{\Oxford}{Subdepartment of Particle Physics, University of Oxford, Oxford OX1 3RH, United Kingdom}
\newcommand{\PennState}{Department of Physics, Pennsylvania State University, State College, Pennsylvania 16802, USA}
\newcommand{\PennU}{Department of Physics and Astronomy, University of Pennsylvania, Philadelphia, Pennsylvania 19104, USA}
\newcommand{\Pittsburgh}{Department of Physics and Astronomy, University of Pittsburgh, Pittsburgh, Pennsylvania 15260, USA}
\newcommand{\IHEP}{Institute for High Energy Physics, Protvino, Moscow Region RU-140284, Russia}
\newcommand{\Rochester}{Department of Physics and Astronomy, University of Rochester, New York 14627 USA}
\newcommand{\RoyalH}{Physics Department, Royal Holloway, University of London, Egham, Surrey, TW20 0EX, United Kingdom}
\newcommand{\Carolina}{Department of Physics and Astronomy, University of South Carolina, Columbia, South Carolina 29208, USA}
\newcommand{\SLAC}{Stanford Linear Accelerator Center, Stanford, California 94309, USA}
\newcommand{\Stanford}{Department of Physics, Stanford University, Stanford, California 94305, USA}
\newcommand{\StJohnFisher}{Physics Department, St. John Fisher College, Rochester, New York 14618 USA}
\newcommand{\Sussex}{Department of Physics and Astronomy, University of Sussex, Falmer, Brighton BN1 9QH, United Kingdom}
\newcommand{\TexasAM}{Physics Department, Texas A\&M University, College Station, Texas 77843, USA}
\newcommand{\Texas}{Department of Physics, University of Texas at Austin, 1 University Station C1600, Austin, Texas 78712, USA}
\newcommand{\TechX}{Tech-X Corporation, Boulder, Colorado 80303, USA}
\newcommand{\Tufts}{Physics Department, Tufts University, Medford, Massachusetts 02155, USA}
\newcommand{\UNICAMP}{Universidade Estadual de Campinas, IFGW-UNICAMP, CP 6165, 13083-970, Campinas, SP, Brazil}
\newcommand{\UFG}{Instituto de F\'{i}sica, Universidade Federal de Goi\'{a}s, CP 131, 74001-970, Goi\^{a}nia, GO, Brazil}
\newcommand{\USP}{Instituto de F\'{i}sica, Universidade de S\~{a}o Paulo,  CP 66318, 05315-970, S\~{a}o Paulo, SP, Brazil}
\newcommand{\Warsaw}{Department of Physics, University of Warsaw, Ho\.{z}a 69, PL-00-681 Warsaw, Poland}
\newcommand{\Washington}{Physics Department, Western Washington University, Bellingham, Washington 98225, USA}
\newcommand{\WandM}{Department of Physics, College of William \& Mary, Williamsburg, Virginia 23187, USA}
\newcommand{\Wisconsin}{Physics Department, University of Wisconsin, Madison, Wisconsin 53706, USA}
\newcommand{\deceased}{Deceased.}

\affiliation{\ANL}
\affiliation{\Athens}
\affiliation{\BNL}
\affiliation{\Caltech}
\affiliation{\Cambridge}
\affiliation{\UNICAMP}
\affiliation{\FNAL}
\affiliation{\UFG}
\affiliation{\Harvard}
\affiliation{\HolyCross}
\affiliation{\Houston}
\affiliation{\IIT}
\affiliation{\Indiana}
\affiliation{\Iowa}
\affiliation{\UCL}
\affiliation{\Minnesota}
\affiliation{\Duluth}
\affiliation{\Otterbein}
\affiliation{\Oxford}
\affiliation{\Pittsburgh}
\affiliation{\RAL}
\affiliation{\USP}
\affiliation{\Carolina}
\affiliation{\Stanford}
\affiliation{\Sussex}
\affiliation{\TexasAM}
\affiliation{\Texas}
\affiliation{\Tufts}
\affiliation{\Warsaw}
\affiliation{\WandM}

\author{P.~Adamson}
\affiliation{\FNAL}







\author{C.~Backhouse}
\affiliation{\Oxford}




\author{G.~Barr}
\affiliation{\Oxford}









\author{M.~Bishai}
\affiliation{\BNL}

\author{A.~S.~T.~Blake}
\affiliation{\Cambridge}


\author{G.~J.~Bock}
\affiliation{\FNAL}

\author{D.~J.~Boehnlein}
\affiliation{\FNAL}

\author{D.~Bogert}
\affiliation{\FNAL}




\author{S.~V.~Cao}
\affiliation{\Texas}


\author{J.~D.~Chapman}
\affiliation{\Cambridge}


% ...

\collaboration{The MINOS Collaboration}
\noaffiliation

\date{\today}

\begin{abstract}
This paper reports measurements of atmospheric neutrino
and antineutrino interactions in the MINOS Far Detector, 
based on 2553~live-days (37.9 kton-years) of data. 
A total of 2072~candidate events are observed. 
These are separated into 905~contained-vertex muons 
and 466~neutrino-induced rock-muons, both produced by
charged-current $\nu_{\mu}$ and $\overline{\nu}_{\mu}$ interactions,
and 701~contained-vertex showers, composed mainly
of charged-current $\nu_{e}$ and $\overline{\nu}_{e}$ interactions
and neutral-current interactions.
The curvature of muon tracks in the
magnetic field of the MINOS Far Detector is used to select
separate samples of $\nu_{\mu}$ and $\overline{\nu}_{\mu}$ events.
The observed ratio of $\overline{\nu}_{\mu}$ to $\nu_{\mu}$ events
is compared with the Monte Carlo simulation,
giving a double ratio of
 $R^{data}_{\overline{\nu}/\nu}/R^{MC}_{\overline{\nu}/\nu}
  = 1.03 \pm 0.08 \,(\mbox{stat.}) \pm 0.08\,(\mbox{syst.})$.
The $\nu_{\mu}$ and $\overline{\nu}_{\mu}$ data are separated into bins of $L/E$ resolution,
based on the reconstructed energy and direction of each event,
and a maximum likelihood fit to the observed
$L/E$ distributions is used to determine the 
atmospheric neutrino oscillation parameters.
This fit returns 90\% confidence limits of 
$|\Delta m^{2}| = (1.9 \pm 0.4 ) \times 10^{-3} \mbox{\,eV}^{2}$
and $\mbox{sin}^{2} 2\theta > 0.86$.
The fit is extended to incorporate separate 
$\nu_{\mu}$ and $\overline{\nu}_{\mu}$ oscillation parameters,
returning 90\% confidence limits of
$|\Delta m^{2}|-|\Delta \overline{m}^{2}| = 0.6^{+2.4}_{-0.8} \times 10^{-3} \mbox{\,eV}^{2}$
on the difference between the squared-mass splittings for
neutrinos and antineutrinos.
\end{abstract}


\maketitle

\section{\label{Introduction}Introduction}

It has now been firmly established by experiment
that muon neutrinos produced by cosmic-ray showers 
in the atmosphere undergo oscillations.
The data are well described 
by $\nu_{\mu} \rightarrow \nu_{\tau}$ neutrino oscillations, 
and measurements of the oscillation parameters have been 
made by Super-Kamiokande (SK)~\cite{superk1,superk2,superk3},
MACRO~\cite{macro}, Soudan~2~\cite{soudan2} and MINOS~\cite{minoscontainedvertex,minosupwardmuon}. 
The atmospheric neutrino results are strongly supported by
long-baseline experiments, which observe corresponding 
oscillations in accelerator beams of muon neutrinos.
Beam neutrino measurements have been made by
K2K~\cite{k2k}, T2K~\cite{t2k} and MINOS~\cite{minos1,minos2,minos3},
The MINOS beam data analysis, which uses a two-flavor model of neutrino oscillations,
returns best fit values of 
$|\Delta m^{2}| = (2.32^{+0.12}_{-0.08}) \times10^{-3} \mbox{\,eV}^{2}$
and $\mbox{sin}^2 2\theta = 1.00_{-0.06}$ for the oscillation parameters~\cite{minos3}.

The MINOS experiment has performed separate measurements of
antineutrino oscillations~\cite{minosnubar,minosrhc1,minosrhc2}
by identifying antineutrino interactions in the Fermilab NuMI accelerator beam~\cite{numibeam}.
A precision measurement of these oscillations
has been made by operating the NuMI beam in a $\overline{\nu}_{\mu}$-enhanced configuration.
Using the $\overline{\nu}_{\mu}$-enhanced data set, 
the $\overline{\nu}_{\mu}$ oscillation parameters are measured to be
$|\Delta \overline{m}^{2}| = [2.62^{+0.31}_{-0.28}\,(\mbox{stat.})\pm0.09\,(\mbox{syst.})] \times10^{-3} \mbox{\,eV}^{2}$
and $\mbox{sin}^2 2\overline{\theta} = 0.95^{+0.10}_{-0.11}\,(\mbox{stat.})\pm0.01\,(\mbox{syst.})$~\cite{minosrhc2}.
Such studies are of interest, as an apparent difference between 
the $\nu_{\mu}$ and $\overline{\nu}_{\mu}$ oscillation parameters 
could indicate new physics. 
In particular, separate $\nu_{\mu}$ and $\overline{\nu}_{\mu}$ measurements
can be used to study models of non-standard neutrino interactions~\cite{nsi1,nsi2},
and probe CPT symmetry in the neutrino sector~\cite{cpt1,cpt2}.

The SK experiment has also studied oscillations in atmospheric neutrinos
and antineutrinos. 
Although SK cannot distinguish $\nu_{\mu}$ from $\overline{\nu}_{\mu}$
on an event-by-event basis, they have performed a statistical analysis 
of their data and the results are consistent with equal $\nu_{\mu}$ and $\overline{\nu}_{\mu}$
oscillation parameters~\cite{sknubar}.

The MINOS experiment is able to study atmospheric neutrinos 
and antineutrinos separately using its 5.4\,kton Far Detector, 
which is located 705\,m underground (2070\,m water-equivalent)
in the Soudan mine, Minnesota.
At this depth, the incident flux of cosmic-ray muons is reduced 
by a factor of $10^{6}$ relative to the surface.
By applying a series of selection requirements,
the cosmic-ray muon background can be reduced by a 
further factor of $10^{6}$, 
yielding a clean sample of atmospheric neutrino signal events.
The MINOS Far Detector is magnetized, which
enables atmospheric $\nu_{\mu} + N \rightarrow \mu^{-} + X$ 
and $\overline{\nu}_{\mu} + N \rightarrow \mu^{+} + X$ 
charged-current (CC) interactions to be separated based 
on the curvature of the muons.

MINOS has been collecting atmospheric neutrino data since~2003 
and has previously published charge-separated analyses of 
contained-vertex muons~\cite{minoscontainedvertex} 
and neutrino-induced rock-muons~\cite{minosupwardmuon,rebel}, 
based on 418 and 854~live-days of data, respectively. 
Both samples are largely composed of atmospheric neutrino
$\nu_{\mu}$ and $\overline{\nu}_{\mu}$ CC interactions,
which are identified by the presence of a primary muon track
in the reconstructed event.
For contained-vertex muons, the reconstructed interaction vertex 
is contained inside the fiducial volume of the detector. 
The sample includes both fully-contained muons, which stop in the detector, 
and partially-contained muons, which exit the detector.
The muon track is typically accompanied by some vertex shower activity,
generated by the hadronic system,
which is used to fully reconstruct the neutrino energy.
For neutrino-induced rock-muons, the reconstructed vertex is 
outside the fiducial volume. The selected muons enter the detector 
in an upward-going or horizontal direction and can be either 
stopping or through-going.

The atmospheric neutrino analysis presented here 
is based on an updated data set of 2553~live-days, 
collected between August 2003 and March 2011. 
The contained-vertex muon and neutrino-induced rock-muon 
samples have been combined into a single analysis, 
along with an additional sample of contained-vertex showering neutrinos, 
which are mainly composed of $\nu_{e}$ and $\overline{\nu}_{e}$ CC interactions
and neutral-current (NC) interactions.
The data are compared to the hypothesis of 
$\nu_{\mu} \rightarrow \nu_{\tau}$ and
$\overline{\nu}_{\mu} \rightarrow \overline{\nu}_{\tau}$
two-flavor vacuum oscillations.

\section{\label{Detector}The MINOS Far Detector}

The MINOS Far Detector~\cite{minosdetector} is a 
steel-scintillator calorimeter, containing 486~octagonal planes 
of 2.54\,cm thick steel, interleaved with planes of
1\,cm thick extruded polystyrene scintillator
and air gaps of 2.4\,cm thickness.
The planes are vertical, with a height of 8\,m.
Each scintillator plane is divided into 
192 strips of width 4.1\,cm, aligned at $\pm45$~degrees 
to vertical. The direction of the strips alternates 
from plane to plane. 
The scintillation light is collected using wavelength-shifting fibers,
which are embedded in the strips. At the ends of each strip,
the emitted light is transported by clear optical fibers
to multi-anode photo-multiplier tubes.

The detector comprises two super-modules, 
of length 14.8\,m and 14.0\,m, separated by a gap of 1.1\,m. 
Each super-module is magnetized toroidally to an
average field of 1.3\,T using a current loop that 
runs through a 25\,cm diameter coil hole along 
the central axis of the super-module and then 
returns below the super-module. 
The MINOS coordinate system is right-handed, with the $y$-axis
pointed vertically upwards and the $z$-axis directed horizontally
along the central axis of the detector, such that 
beam neutrinos have a forward-going $z$-direction.
The directions of the scintillator strips define a pair of
diagonal axes $U=(x+y)/\surd 2$ and $V=(-x+y)/\surd 2$. 
Each strip provides a 2D spatial point
in either the $U-z$ or $V-z$ coordinate systems,
denoted the $U$ and $V$ views, respectively.

The vertical alignment of the planes presents a 
source of difficulty in separating contained-vertex 
atmospheric neutrinos from the cosmic-ray muon
background. Steep cosmic-ray muons incident 
on the detector between two planes can travel a 
significant distance into the detector before 
entering the scintillator, and therefore appear 
as contained-vertex atmospheric neutrino events. 
To reduce the background, a scintillator veto shield 
has been constructed above the detector, and is used 
to tag cosmic-ray muons entering the detector. 
The veto scintillator modules are grouped into four 
sections, two per super-module, with a double layer 
on the top surface of the detector, and single layers 
diagonally above and at each side of the detector. 
To prevent gaps, adjacent modules overlap each other.
The majority of cosmic-ray muons pass through two 
layers of scintillator before entering the detector
and can therefore be vetoed with high efficiency. 

The veto shield is used to reject cosmic-ray muon background
in the selection of contained-vertex tracks and showers.
An event is rejected if any activity is observed 
in the section of shield above the event vertex 
within a time window of $\pm 50$\,ns. 
The shield efficiency is determined using samples
of cosmic-ray muons from across the entire data set. 
The fraction of cosmic-ray muons vetoed by 
the shield is measured to be $96.6$\%$\pm 0.3$\%\,$(\mbox{syst.})$.
The systematic uncertainty is obtained by modifying 
the criteria used to select the cosmic-ray muon samples
and calculating the resulting variation in shield efficiency.

At each stage of the contained-vertex event selection, 
the cosmic-ray muon background predictions are 
derived directly from the data by scaling down 
the observed distributions of vetoed events
according to the measured shield efficiency. 
A small fraction of the atmospheric neutrino signal 
is also vetoed as a result of accidental coincidence 
with noise in the shield. The loss of signal is determined 
to be $1.0$\%$\pm 0.2$\%\,$(\mbox{syst.})$, found by overlaying samples 
of veto shield data on simulated atmospheric neutrino events. 
The systematic uncertainty reflects the
time-dependent variations in the veto shield data rates.

Since March 2005, the MINOS Far Detector has been used 
to study neutrino interactions from the Fermilab NuMI accelerator beam.
The beam neutrinos are identified in the data
by searching in 100\,$\mu$s time windows, 
extrapolated from the beam spill times.
These windows, which correspond to approximately 
0.01\% of the Far Detector live time, 
are removed from the atmospheric neutrino analysis. 
For the majority of running, the detector has been
magnetized to focus forward-going negatively 
charged muons from beam neutrino interactions. 
However, the magnetic field was reversed during 
the $\overline{\nu}_{\mu}$-enhanced beam running, 
and also for a period of several months prior to 
beam start-up, for the purpose of studying the 
cosmic-ray muon charge ratio~\cite{minoschargeratio}.

Only data collected with both the main detector and 
veto shield fully operational are used in the analysis. 
The final data set corresponds to 2553 live-days,
an exposure of 37.9 kton-years.

\section{\label{MonteCarlo}Atmospheric neutrino Monte Carlo simulation}

The MINOS Monte Carlo simulation uses separate
programs to generate contained-vertex atmospheric neutrino 
interactions inside the Far Detector and neutrino-induced muons
from interactions in the surrounding rock.
For contained-vertex atmospheric neutrino interactions,
the {\sc neugen3} simulation~\cite{neugen3} is used to generate 
the interactions and hadronic final states. 
The transport of hadronic particles is then modelled using the 
{\sc gcalor} simulation~\cite{gcalor}.
For neutrino-induced rock-muons, 
the {\sc nuance} generator~\cite{nuance} is used,
with the {\sc grv94} parton distribution functions~\cite{grv94}.
The {\sc nuance} simulation then propagates the muons 
from the rock to the edges of the detector.
For both Monte Carlo samples, 
a {\sc geant3}~\cite{geant3} simulation of the Far Detector
is used to model particle transport and detector response.

For contained-vertex atmospheric neutrino events, 
the simulation uses the flux calculation
of Barr {\it et al.}~\cite{bartol3D} (Bartol 3D).
For neutrino energies below 10\,GeV, 
this calculation is based on a 3D simulation,
with separate flux tables provided for 
the cases of solar minimum and solar maximum.
Above 10\,GeV, a 1D simulation has been used
and a single set of flux tables is provided.
For neutrino-induced rock-muons, 
where the majority of parent neutrinos have energies greater than 10\,GeV, 
the simulation uses the earlier 1D calculation 
by the Bartol group~\cite{bartol96} (Bartol 1D). 
The rock-muons are then reweighted as a function 
of their parent neutrino energy and zenith angle 
using the ratio of the Bartol~3D and Bartol~1D fluxes.
Both the contained-vertex and rock-muon samples
use distributions of neutrino production height
obtained from a separate simulation of cosmic-ray interactions 
in the atmosphere and parameterized  
in terms of neutrino energy and zenith angle~\cite{soudan2}.

The Far Detector data set spans a significant fraction
of the full solar cycle. For atmospheric neutrinos above 500\,MeV
(the MINOS energy threshold),
the predicted neutrino interaction rate is 7\% higher at
solar minimum than solar maximum. 
These solar cycle effects are accounted for by
taking a weighted average of the Bartol 3D fluxes
calculated at solar minimum and solar maximum.
The variation of solar activity over time is determined 
by parameterizing available atmospheric neutron data 
from the CLIMAX experiment~\cite{climax}. 
Combining these data with the Far Detector running periods,
the fluxes at solar minimum and solar maximum 
are combined in proportions 70\% and 30\%, respectively. 

\begin{figure}[!tb]
   \includegraphics[width=\columnwidth]{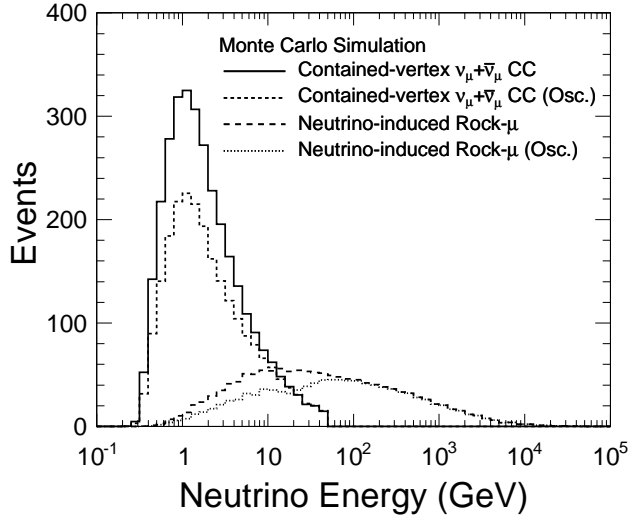}
   \caption{\label{fig_spectrum}
     The simulated atmospheric neutrino energy spectrum
     in the MINOS Far Detector for 2553 live-days of data.
     Separate distributions are plotted for contained-vertex
     neutrino interactions and neutrino-induced rock-muons
     showing the predictions for the case of no oscillations, 
     and for oscillations with 
     $\Delta m^{2} = 2.32 \times 10^{-3} \mbox{\,eV}^{2}$ 
     and $\mbox{sin}^{2} 2\theta = 1.0$. 
     The contained-vertex neutrino interactions are generated 
     in the range $0.2-50$\,GeV, with a median value of 2\,GeV;
     the neutrino-induced rock-muons range up to 
     neutrino energies of 10\,TeV, with a median value of 50\,GeV. 
     The effect of $\nu_{\mu} \rightarrow \nu_{\tau}$ oscillations
     is visible for neutrino energies below 100\,GeV.}
 \end{figure}

Figure~\ref{fig_spectrum} shows the simulated 
atmospheric neutrino energy spectrum for 
contained-vertex neutrino interactions and neutrino-induced rock-muons, 
plotted for the case of no oscillations, 
and for oscillation parameters of 
$\Delta m^{2} = 2.32\times10^{-3} \mbox{\,eV}^{2}$ 
and $\mbox{sin}^2 2\theta = 1.0$~\cite{minos3}.
These oscillation parameters are also used 
to calculate predicted atmospheric neutrino event rates
in Sections~\ref{Selection} and \ref{ChargeRatio} of this paper.

 %
 %

\subsection{Systematic uncertainties in atmospheric neutrino simulation}

The predicted atmospheric neutrino event rates have
large uncertainties arising from the atmospheric neutrino
flux and interaction models. 
The Bartol group has carried out a detailed study of the
systematic uncertainties in their 3D flux model~\cite{bartolfluxerrors}.
These flux uncertainties have also been studied 
by comparing the Bartol model with the alternative 
3D calculations of Battistoni {\it et al.}~\cite{battistoni}
and Honda {\it et al.}~\cite{honda}.
The systematic uncertainties used in this analysis 
are based on the results of the Bartol study 
but are also found to cover the differences
between the different flux models.

The dominant source of uncertainty in the overall 
rate of contained-vertex neutrinos and 
neutrino-induced rock-muons is the 
systematic uncertainty in the normalization
of the atmospheric neutrino flux simulation.
This overall uncertainty increases with neutrino energy 
due to rising uncertainties in the 
primary cosmic-ray flux and hadroproduction models.
For contained-vertex neutrinos,
which have a median energy of 2\,GeV
and lie primarily below $10$\,GeV,
an overall uncertainty of $15$\% is applied in this analysis.
For neutrino-induced rock-muons, 
where the parent neutrino has a median energy of 50\,GeV,
with an energy spectrum that ranges up to $10$\,TeV, 
a larger uncertainty of $25\%$ is applied.

At low neutrino energies, many of the systematic uncertainties 
in the flux model cancel in the ratios of different flux components.
In the $1-5$\,GeV region, the uncertainties in the
($\nu_{\mu}$+$\overline{\nu}_{\mu}$)/($\nu_{e}$+$\overline{\nu}_{e}$)
and $\nu_{\mu}/\overline{\nu}_{\mu}$ flux ratios,
and in the up-down ratio of upward-going neutrinos to downward-going neutrinos,
are calculated to be smaller than 5\%~\cite{bartolfluxerrors}. 
For the analysis presented here,
the uncertainty in the $\nu_{\mu}/\overline{\nu}_{\mu}$ 
ratio is of greatest importance. At energies below 10\,GeV,
cosmic-ray hadroproduction predominantly yields pions, 
with each charged pion producing a single pair of $\nu_{\mu}$ and $\overline{\nu}_{\mu}$ 
in its decay chain. Therefore, the $\nu_{\mu}/\overline{\nu}_{\mu}$ 
ratio approaches unity, with a high degree of cancellation 
in its systematic uncertainty. For the contained-vertex
atmospheric neutrino sample, a conservative uncertainty 
of 4\% is placed on this ratio. At energies above 10\,GeV,
the cancellations in the ratio quickly diminish, 
as the large uncertainty in the kaon component of hadroproduction 
becomes a significant factor, and an increasing fraction of 
atmospheric muons strike the ground before decaying.
Therefore, for the neutrino-induced rock-muon sample, 
a larger uncertainty of $10$\% is placed on the $\nu_{\mu}/\overline{\nu}_{\mu}$ ratio.

Additional systematic uncertainties in the predicted 
atmospheric neutrino event rate arise from the neutrino 
interaction model.
The uncertainty in the total $\nu_{\mu}$ CC cross-section 
peaks at $8$\% in the $1-5$\,GeV region~\cite{minosprd}, 
corresponding to the transition region between the models 
of quasi-elastic and resonance neutrino interactions. 
At higher energies, where deep inelastic interactions are dominant,
the total interaction cross-section is well constrained 
by experiment, and the uncertainty falls to $2$\%.
For antineutrinos, where there is limited experimental data below $5$\,GeV,
the predicted $\overline{\nu}_{\mu}$ CC cross-section 
has a larger uncertainty. For the analysis presented here,
an energy-dependent systematic uncertainty band
on the $\nu_{\mu}/\overline{\nu}_{\mu}$
cross-section ratio has been calculated by
varying the input parameters to the {\sc neugen3} interaction model 
according to their given uncertainties~\cite{neugensys}.
The average uncertainty in the $\nu_{\mu}/\overline{\nu}_{\mu}$ cross-section ratio
is then calculated by integrating across this uncertainty band, 
weighting each bin of neutrino energy by the predicted rate 
of atmospheric neutrino $\nu_{\mu}$ and $\overline{\nu}_{\mu}$ CC interactions.
For the contained-vertex neutrino sample, 
this procedure yields an overall uncertainty of 8.5\%. 
For the neutrino-induced rock-muon sample,
where the majority of events are produced by 
deep inelastic neutrino interactions,
the calculation returns a smaller uncertainty of 4\%.

The atmospheric neutrino event rate at the Soudan mine 
has been previously measured by the Soudan~2 experiment.
The Soudan~2 analysis of atmospheric electron neutrinos
indicates that the predicted interaction rates obtained by combining 
the Bartol 3D flux model and {\sc neugen} cross-section model 
should be scaled by $0.91 \pm 0.07$~\cite{soudan2}. However,
although the Soudan~2 and MINOS detectors are located at the
same site, Soudan~2 has a lower neutrino energy threshold of
300\,MeV, compared with 500\,MeV for MINOS. 
An analysis of contained-vertex showers from atmospheric neutrinos 
by MINOS, based on 418 live-days of data, yields a scale factor of 
$1.08 \pm 0.12\,(\mbox{stat.}) \pm 0.08\,(\mbox{syst.})$~\cite{speakman}.
The systematic uncertainties applied in this analysis 
cover both these measurements.

\section{\label{Reconstruction}Event reconstruction}

The data are reconstructed using an algorithm 
which identifies the track and shower topologies in 
each event~\cite{blake}. Reconstructed tracks typically 
contain hits in one strip per plane and are principally produced 
by muons; reconstructed showers contain hits in multiple strips 
per plane and are produced by hadronic and 
electromagnetic particles.

Initially, the particle tracks and showers are reconstructed 
independently in each of the $U$ and $V$ views; 
these 2D views are then matched to generate a 3D event.
A Kalman filter algorithm is used to determine the trajectory
of each muon track, accounting for energy loss in the detector
and curvature in the magnetic field~\cite{marshall}.
This algorithm also reconstructs the start and end points of each track,
which are distinguished using timing information.
For tracks where the end point lies inside the
fiducial volume, the muon momentum is reconstructed 
from the measured track length; for exiting tracks,
the momentum is obtained from the fitted
track curvature. In both cases, the fitted curvature 
is used to determine the muon charge sign.
For atmospheric neutrino events containing a reconstructed track,
the interaction vertex is given by the start point of the track; 
if there is only a reconstructed shower, the vertex is given by 
the centroid of the shower.

The propagation direction of each muon along its reconstructed track
is determined using timing information.
The MINOS Far Detector has a single-hit timing resolution of 
approximately $2.5$\,ns, which enables the muon
direction to be reconstructed with high purity for
tracks spanning ten or more scintillator planes. 
The Far Detector timing system is calibrated using 
cosmic-ray muons, which are used to determine the 
time offsets in each readout channel and to correct 
for shifts in these offsets resulting from swapped
readout components~\cite{blake}.

The detector is calibrated using a combination of 
LED light injection and the average pulse height response 
of each strip using cosmic ray muons~\cite{minosdetector}. 
A minimum-ionizing muon passing through a scintillator strip 
at normal incidence generates a combined signal of 
approximately 10 photo-electrons (PEs). 
The selection of contained-vertex tracks and showers 
makes use of the energy profile of events.

For contained-vertex atmospheric $\nu_{\mu}$ and $\overline{\nu}_{\mu}$ events, 
the emitted muon is typically accompanied by some 
reconstructed shower activity at the interaction vertex, 
produced by the hadronic system.
The total hadronic energy is determined by summing 
the calibrated pulse heights in the reconstructed shower. 
For low energy showers, large fluctuations can occur,
degrading the hadronic energy resolution. 
To reduce the size of these fluctuations, 
the pulse heights are first raised to a power
before being summed together.
The exponent used in this procedure 
is increased as a function of shower energy
from a minimum of 0.25 at the lowest shower energies 
to a maximum of 1.0 for shower energies above 18\,GeV~\cite{culling}.
Studies of simulated atmospheric neutrinos show that,
relative to a linear summation of pulse heights, 
the hadronic energy resolution improves from 55\% to 45\% 
for reconstructed showers in the 1\,GeV region.

\section{\label{Selection}Event selection}

An initial selection is applied to all events, 
ensuring a good reconstruction quality. 
The selected events are then separated into a 
track-like sample containing reconstructed tracks 
that span $8$~or more planes, and a shower-like sample 
containing reconstructed showers that span $4$~or more planes. 
The track-like sample is used for the selection 
of contained-vertex muons and neutrino-induced rock-muons; 
the shower-like sample is used for the selection of contained-vertex showers. 
Initially, 2\% of events are placed in both the track-like and shower-like samples.
Any duplicate events are removed from the shower-like 
sample after the full selection has been applied.
After the initial selection, the observed event rate 
is 55,000~events/day, dominated by cosmic-ray muons.
The corresponding predicted atmospheric neutrino rates are 
0.8~events/day from contained-vertex interactions, 
and 0.3~events/day from neutrino-induced rock-muons,
after accounting for oscillations.

The atmospheric neutrino signal is separated
from the cosmic-ray background using two
characteristic signatures of atmospheric
neutrino interactions: either a reconstructed
vertex inside the fiducial volume; 
or a reconstructed upward-going or horizontal muon trajectory.
A set of requirements on event containment and topology 
is first applied to select contained-vertex tracks and showers, 
using the veto shield to reduce the cosmic-ray muon background. 
A set of requirements on event timing information and length 
is then applied to select upward-going and horizontal muons 
produced by neutrino interactions in the detector or surrounding rock. 
The full selection is described in the
following sections.

  \begin{figure}[!tb]
   \includegraphics[width=\columnwidth]{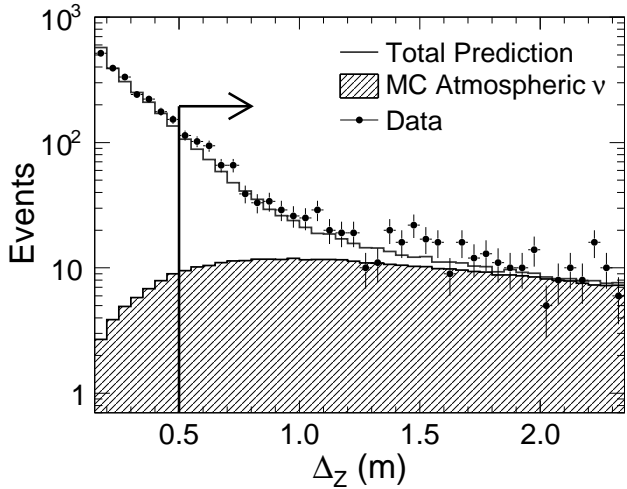}
   \caption{\label{fig_cv_trace_cut}
     Distributions of the trace variable, $\Delta_{Z}$, 
     for contained-vertex tracks.
     This estimates the distance in $z$ travelled by a
     cosmic-ray muon inside the detector before first entering 
     the scintillator.
     The hatched histogram shows the simulated prediction 
     for the atmospheric neutrino signal; the solid line shows 
     the predicted total rate, given by the sum of the signal 
     and the cosmic-ray muon background; the points show
     the observed data. 
     The background distribution is peaked towards low values 
     of $\Delta_{Z}$, and the arrow indicates the selection 
     applied to reduce the background.}
 \end{figure}

 \begin{figure*}[!tb]
   \includegraphics[width=\textwidth]{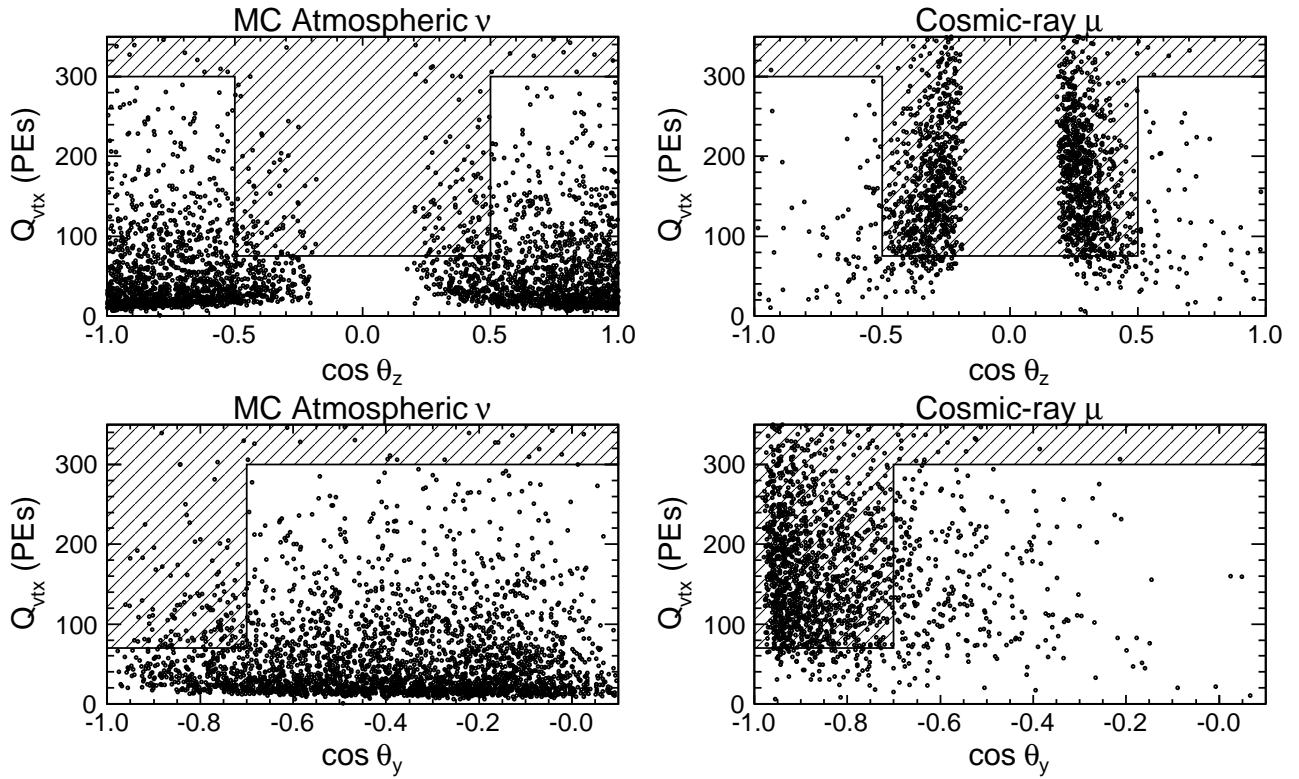}
   \caption{\label{fig_cv_topology_cut} 
     Distributions of the pulse height at the upper end of the track
     ($Q_{vtx}$), plotted against the z-component ($\cos\theta_{z}$) 
     and y-component ($\cos\theta_{y}$) of the downward track direction.
     The distributions are plotted for contained-vertex muons
     that pass the trace and topology requirements and span fewer than 25 planes. 
     The plots on the left show simulated atmospheric neutrinos;
     those on the right show the cosmic-ray muon background.
     The background events are associated with large pulse heights 
     and directions parallel to the vertically-aligned scintillator planes.
     The hatched area is the region rejected by the pulse height and direction 
     selection criteria as part of the topology requirements.
     Note that the requirement of a track spanning $\geq 8$ planes
     causes the acceptance to drop to zero as $|\cos\theta_{z}|$ approaches 0,
     and as $\cos\theta_{y}$ approaches $-1$.}
 \end{figure*}

\subsection{Selection of contained-vertex tracks}

The contained-vertex track selection criteria identify
$\nu_{\mu}$ and $\overline{\nu}_{\mu}$ CC atmospheric neutrinos.
For this sample,
the veto shield selection is first used to reduce the level 
of cosmic-ray muon background. A set of containment and 
topology selection criteria are then applied to the remaining tracks. 
For each track, the reconstructed trajectory has two ends,
corresponding to the first and last scintillator hits
on the track.
Since cosmic-ray muons are incident from above,
the majority of these selection criteria are applied at the 
upper end of the track. 
The following selection criteria are applied \cite{howcroft,blake,chapman}:

\begin{enumerate}

 \item \textit{Fiducial Cuts:} 
   The reconstructed track vertex is required to lie 
   within a fiducial volume starting $0.2$\,m inside any edge 
   of the detector, $5$ planes from the ends of each super-module, 
   and $0.4$\,m from the center of the coil hole. 
   To reject cosmic-ray muons that enter the detector 
   through the coil hole, the coil cut is increased to 1\,m 
   in the first and last 20 planes of the detector.
   In addition, if the vertex is reconstructed at 
   the lower end of the track from timing information, 
   these selection criteria are also applied at the 
   upper end of the track.

 \item \textit{Trace Cut:}
   The cosmic-ray muons that pass the fiducial requirements
   typically enter the detector at a small angle
   to the planes, and can travel a significant distance
   through the detector before entering the scintillator.
   However, the distance travelled along the $z$-axis is typically small, 
   and so the background can be reduced by placing a
   minimum requirement on this distance.
   For each event, 
   a detector entry point is estimated by extending 
   the reconstructed trajectory at the upper end 
   of the track upwards to the edge of the detector.
   The displacement along the $z$-axis, $\Delta_{Z}$,
   between the entry point and upper end
   of the track is then calculated 
   (this quantity is referred to as the ``trace'').
   Figure~\ref{fig_cv_trace_cut} shows the predicted
   and observed distributions of the trace variable. 
   The cosmic-ray muon background is
   peaked towards low values, whereas the 
   atmospheric neutrino signal distribution has a 
   flatter distribution. To reduce the background,
   events are required to satisfy $\Delta_{Z} > 0.5$\,m.
  
 \item \textit{Topology Cuts:} 
   The cosmic-ray muon background events that pass 
   the trace cut typically travel a significant distance 
   in a single steel plane and its associated air gap 
   before entering the scintillator. 
   A~number of these cosmic-ray muons undergo significant 
   bending in the magnetic field, and some muons reverse 
   the horizontal component of their direction.  
   As a result, the reconstruction may miss the first hit
   on the track or underestimate the steepness of the track.
   These events are characterized by clusters of hits
   above the upper end of the reconstructed track. 
   To reduce this background, 
   the charge-weighted mean and RMS displacements of strips,
   denoted
   $\langle \Delta_{UV} \rangle$ and $\langle \Delta^{2}_{UV}\rangle^{\frac{1}{2}}$,
   are calculated separately in the $U$ and $V$ views
   for a $\pm 4$~plane window around the upper end of the track.
   Events are rejected if $\langle \Delta_{UV} \rangle > 0.25$\,m,
   indicating that significant energy has been deposited 
   above the track,
   or if $\langle \Delta^{2}_{UV} \rangle^{\frac{1}{2}} > 0.5$\,m,
   indicating that there was significant scatter at the upper end of the track. 
   A set of 3D displacements is also calculated 
   using the same $\pm 4$~plane window,
   by combining all possible pairs of $U$ and $V$ strips in adjacent scintillator planes.
   The maximum 3D displacement from the upper end of the track, 
   $\Delta^{max}_{R}$, is then calculated, and events 
   are rejected if $\Delta^{max}_{R} > 1.25$\,m. 

 \item \textit{Pulse Height and Direction Cuts:} 
   The cosmic-ray muon background is also characterized 
   by large deposits of energy at the upper end of the track,
   due to the long distance travelled in the first plane.
   This background is reduced by finding the
   maximum pulse height, $Q_{vtx}$,
   in a $\pm 4$~plane window around the upper end of 
   the track. Events are rejected if $Q_{vtx} > 300$~PEs. 
   The $Q_{vtx}$ requirement is tightened to 75~PEs if
   the track is both short,
   spanning fewer than $25$~planes, and steep, satisfying
   $\cos\theta_{y}>0.7$ or $|\cos\theta_{z}|<0.5$.
   Here, $\theta_{y}$ and $\theta_{z}$ 
   are taken as the angles between the reconstructed
   trajectory at the upper end of the track, and the 
   $y$ and $z$ axes, respectively. Figure~\ref{fig_cv_topology_cut} 
   shows the predicted signal and background distributions of 
   $Q_{vtx}$ as a function of $\theta_{y}$ and $\theta_{z}$ 
   for the short tracks, indicating the selection requirements
   applied to separate the atmospheric neutrino signal from
   the cosmic-ray muon background.

\end{enumerate}

The track containment requirements yield 801 events from the data.
This compares with total predictions of $934 \pm 134$ events 
for no oscillations, 
and $698 \pm 99$ events for oscillations with
$\Delta m^{2} = 2.32 \times 10^{-3} \mbox{\,eV}^{2}$ and $\mbox{sin}^{2} 2\theta = 1.0$.
The uncertainties in these predictions are dominated by the 15\% uncertainty 
in the overall normalization of the contained-vertex Monte Carlo simulation.
The combined $\nu_{\mu}$ and $\overline{\nu}_{\mu}$ CC components
form 92\% of the total predicted event rate before oscillations. 
This component oscillates in the two-flavor model
and therefore represents the signal in the oscillation analysis.
The combined $\nu_{e}$+$\overline{\nu}_{e}$ CC and NC components,
which do not oscillate in the two-flavor model, 
form a 5\% background.
The cosmic-ray muon prediction 
of $34 \pm 3$ events corresponds to a 3\% background level.

\subsection{Selection of upward-going and horizontal tracks}

For upward and horizontal angles, where the 
rock overburden exceeds 14,000\,m water-equivalent, 
the absorption of cosmic-ray muons by the earth is 
sufficiently high that the observed flux of muons 
is dominated by atmospheric muon neutrino interactions~\cite{crouchcurve}.
At the Soudan mine, this corresponds to zenith angles
in the range $\cos\theta_{z}\leq0.14$~\cite{soudan2horizontal}.
Therefore, upward-going and horizontal muons in the 
MINOS Far Detector provide a signature for atmospheric neutrinos.

Upward-going and horizontal muons are selected based on 
the reconstructed zenith angle at the track vertex.
The direction of muon propagation along the track
is reconstructed using timing information. 
This is then used to distinguish between 
the track vertex and end points.
The reconstructed track vertex can either be 
inside or outside the fiducial volume.
Hence, this sample of events provides a source 
of both contained-vertex muons
and neutrino-induced rock-muons.

To determine the track direction from timing information,
two linear fits are applied to the measured 
times of the track hits, as a function of their distance 
along the track. The gradients are constrained to be $\pm 1/c$, 
corresponding to forward and backward propagation along the 
track at the speed of light, $c$. The hits are weighted 
as a function of pulse height to account for the variation 
in the single-hit timing resolution, which is better for 
larger pulse heights due to increased photon statistics. 
For each propagation direction, the goodness of the 
timing fit is given by its RMS timing residual. 
The smaller of the two RMS values is labelled $r_{L}$, 
and the larger is labelled $r_{H}$. The propagation 
direction is determined by the timing fit with the 
smaller RMS residual.

The neutrino-induced muons must be separated from a 
high background of cosmic-ray muons whose direction 
is mis-reconstructed. To ensure that the track direction
is reconstructed unambiguously,
the following selection criteria are applied:

 \begin{figure}[!t]
   \includegraphics[width=\columnwidth]{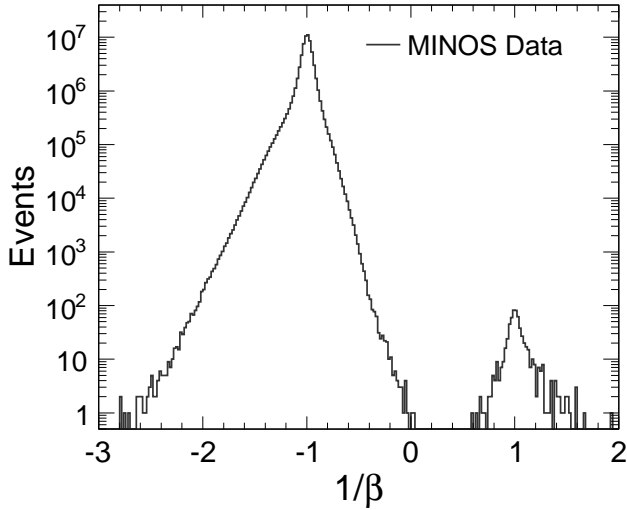} 
   \caption{\label{fig_events_upmu_invbeta}
     Distribution of $1/\beta$ normalized velocity variable, 
     demonstrating the purity of the track direction identification.
     The $1/\beta$ variable is the gradient of a linear fit to 
     the measured times as a function of distance along each track.
     The distribution is plotted for all tracks that pass 
     the topology and timing selections.
     The peak at $-1.0$ corresponds to downward-going muons; 
     the peak at $+1.0$ to upward-going muons. A good separation 
     is achieved between the upward-going neutrino-induced signal, 
     and downward-going cosmic-ray muon background.}
 \end{figure}

 \begin{figure}[!t]
   \includegraphics[width=\columnwidth]{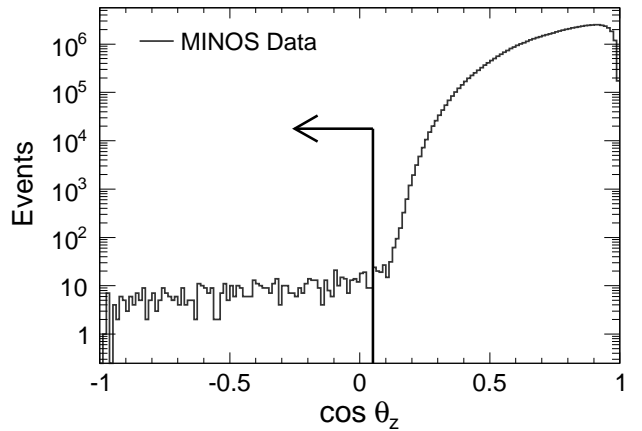} 
   \caption{\label{fig_events_upmu_coszenith}
     Distribution of reconstructed zenith angle for muons
     with good timing and topology. In the range $\cos\theta_{z}>0.10$,
     the observed rate of muons is dominated by the cosmic-ray background
     and falls steeply as the mean rock overburden increases rapidly.
     For $\cos\theta_{z}<0.10$, the distribution flattens,
     as the cosmic-ray muon flux falls below that of neutrino-induced muons.
     To minimize the background from cosmic-ray muons, 
     events are required to satisfy: $\cos\theta_{z}<0.05$. }
 \end{figure}

\begin{enumerate}

 \item \textit{Topology Cuts: }
   To ensure that there are sufficient hits on the track
   for the zenith angle to be reconstructed accurately, 
   and the propagation direction determined 
   unambiguously, reconstructed tracks are required to 
   span more than $15$~planes and to travel more than $1.5$\,m. 
   For upward-going tracks, the track vertex point
   is required to lie below the track end point, or, 
   to account for possible track curvature, no more than 0.5\,m above it. 
   For downward-going tracks, the track vertex requirement 
   is reversed.
   
 \item \textit{Timing Cuts: } 
   To ensure that the muon propagation direction is 
   identified unambiguously from timing information, 
   a set of selection requirements are placed on the 
   quality of the linear timing fits. 
   The difference between the two RMS residuals
   is required to satisfy $r_{L}-r_{H}<-1.66$\,ns,
   a significant fraction of the 
   detector timing resolution.
   In addition, an upper requirement of $r_{L}<4.66$\,ns 
   is placed on the smaller residual 
   and a lower requirement of
   $r_{H}>3.66$\,ns is placed on the larger residual.
   The ratio between the best fit RMS residual, $r_{L}$, 
   and the track length, $l$, also provides a means 
   of selecting events with well-measured timing information.
   Events are required to satisfy $r_{L}/(l/c)<0.577$.

\end{enumerate}
 
These selection criteria identify clean samples 
of both upward-going and downward-going muons 
from the data. As a check on the quality of separation between
these two samples, an additional unconstrained linear timing fit 
is applied to the selected events. The measured times along 
the track are fitted as a function of their upward distance 
along the track. The fits return a gradient, $1/v$,
where $v$ is the reconstructed velocity.
Figure~\ref{fig_events_upmu_invbeta} shows the 
distribution of the normalized gradient, $1/\beta \equiv 1/(v/c)$. 
A~good separation is achieved between the upward-going
events, dominated by neutrino-induced muons,
and the downward-going events, dominated by the
cosmic-ray muon background.

Figure~\ref{fig_events_upmu_coszenith} shows the 
distribution of reconstructed zenith angle 
for the selected events. 
In the region $\cos\theta_{z}>0.10$,
the event rate falls steeply with zenith angle,
as the rapidly increasing rock overburden 
reduces the incident flux of cosmic-ray muons.
In the region $\cos\theta_{z}<0.10$, the event rate
flattens and becomes approximately constant, 
as neutrino-induced muons become the dominant flux component.
A residual background arises from low momentum
cosmic-ray muons that deflect significantly
due to multiple Coulomb scattering in the rock 
and enter the detector in a horizontal direction.
To~minimize the cosmic-ray muon background, 
the selected sample of upward-going and horizontal 
muons is required to satisfy: 
$\cos\theta_{z}<0.05$. An estimate of the remaining
background is obtained from an exponential fit to observed data 
in the region $|\cos\theta_{z}|<0.20$. This method
returns a background prediction of 0.5 selected events.
The cosmic-ray muon background component is neglected
in the subsequent analysis.

Overall, $665$ upward-going and horizontal muons 
are selected from the data.
This compares with total predictions of $882 \pm 146$ events 
in the absence of oscillations, and $623 \pm 113$ events
for input oscillation parameters of $\Delta m^{2} = 2.32 \times 10^{-3} \mbox{\,eV}^{2}$ and $\mbox{sin}^{2} 2\theta = 1.0$,
where the uncertainties are dominated 
by the normalizations of the contained-vertex and 
neutrino-induced rock-muon Monte Carlo simulations.
The selected sample divides into $466$ neutrino-induced rock-muons and $199$ contained-vertex muons.
The latter sample contains $95$ events 
already selected as contained-vertex muons by the containment requirements described above,
along with an additional $104$ upward-going and horizontal events.

\subsection{Selection of contained-vertex showers}

The contained-vertex shower sample is primarily composed
of the $\nu_{e}$+$\overline{\nu}_{e}$ CC and NC atmospheric neutrino component.
For this sample, the main background arises 
from cosmic-ray muons incident at steep angles, which 
radiate large showers and span a small number of planes. 
The veto shield is first applied to reduce
cosmic-ray muon background. The following selection criteria 
are then applied to the remaining showers \cite{speakman}:

 \begin{figure}[!t]
   \includegraphics[width=\columnwidth]{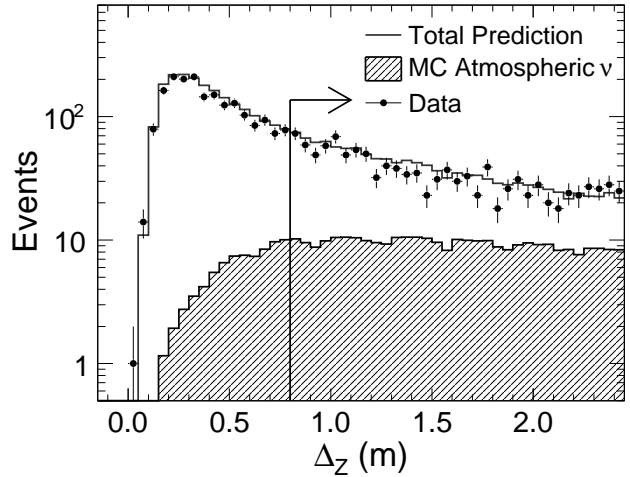}
   \caption{\label{fig_nue_trace_cut}
     Distributions of the trace variable, $\Delta_{Z}$,
     for contained-vertex showers.
     This variable estimates the distance in $z$ travelled by a
     cosmic-ray muon inside the detector before entering the scintillator.
     The hatched histogram shows the simulated prediction 
     for the atmospheric neutrino signal.
     The solid line gives the predicted total rate,
     dominated by the cosmic-ray muon background.
     The points show the observed data. 
     The background is peaked towards low values 
     of $\Delta_{Z}$, 
     since cosmic-ray muons typically travel a small
     distance in $z$ before entering the scintillator.
     The arrow indicates the selection applied on $\Delta_{Z}$ 
     to reduce the background.} 
 \end{figure}   

 \begin{figure*}[!tb]
   \includegraphics[width=\textwidth]{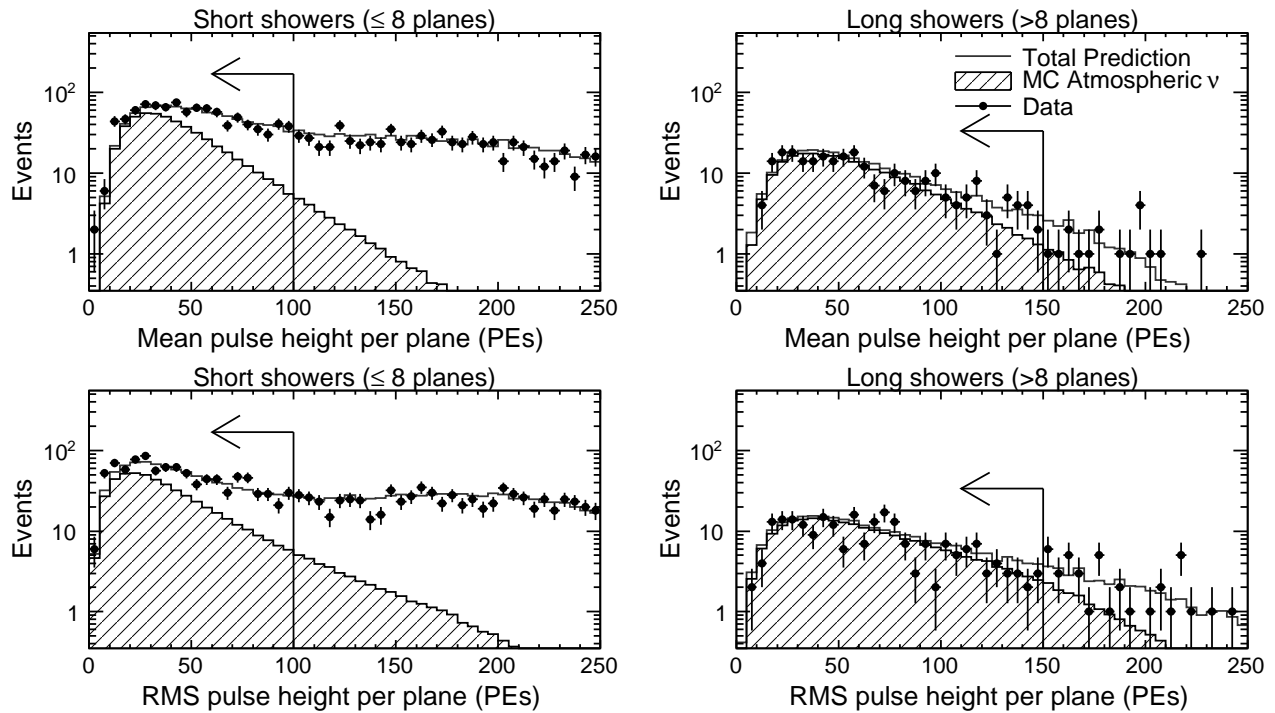}
   \caption{\label{fig_nue_topolgy_cut}
     Distributions of mean and RMS shower pulse height per scintillator plane observed
     for contained-vertex showers that pass the fiducial and trace cuts.
     The hatched histogram shows the simulated prediction 
     for the atmospheric neutrino signal.
     The solid line gives the total prediction,
     dominated by the cosmic-ray muon background.
     The points show the observed data. 
     The arrows indicate the selection applied
     to reduce the background. } 
 \end{figure*}   

 \begin{figure}[!b]
   \includegraphics[width=\columnwidth]{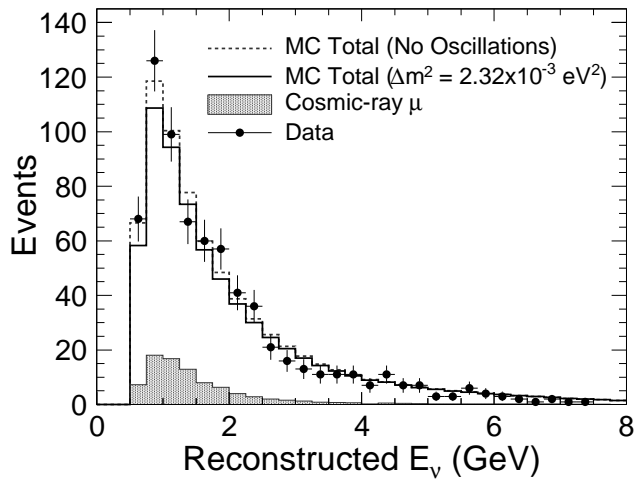}
   \caption{ \label{fig_events_nue_energy}
     Distributions of reconstructed neutrino energy, 
     plotted~for selected contained-vertex showers. 
     The dotted line shows the prediction for no oscillations; 
     the solid line shows the prediction for
     $\Delta m^{2} = 2.32 \times 10^{-3} \mbox{\,eV}^{2}$ 
     and $\mbox{sin}^{2} 2\theta = 1.0$; 
     the shaded histogram shows the predicted cosmic-ray muon background; 
     the points with errors show the observed data.
     The $\nu_{\mu}$+$\overline{\nu}_{\mu}$ CC component is small
     and hence
     the total prediction does not depend strongly on the oscillation parameters.} 
 \end{figure}

\begin{enumerate}

 \item \textit{Fiducial and Trace Cuts: }
   The fiducial requirements described above for contained-vertex tracks
   are also applied in the selection of contained-vertex showers.
   For reconstructed showers, the vertex resolution is poorer 
   and cosmic-ray background level larger.
   Therefore, a tighter fiducial requirement of $0.4$\,m
   is applied at each edge of the detector for this sample. 
   The shower direction also
   has a poorer resolution, and therefore a tighter
   trace cut of $0.8$\,m is applied.
   Figure \ref{fig_nue_trace_cut} shows the 
   predicted and observed $\Delta_{Z}$ distributions 
   for contained-vertex showers. 

 \item \textit{Topology Cuts: } 
   Further selection requirements are applied to identify 
   the characteristic topologies of neutrino-induced showers.
   The longitudinal profile of these showers typically
   rises to a maximum and then falls smoothly,
   whereas showers generated by cosmic-ray muons
   typically deposit a lot of energy in a single plane, 
   or contain large fluctuations between planes. 
   The cosmic-ray background is also found to be higher 
   for shorter showers. Therefore, events are separated 
   into short ($\leq 8$~planes) and long ($>8$~planes) samples, 
   and tighter selection criteria are applied to the short sample.
   To characterize the shower topology,
   the mean and RMS number of strips per plane ($\langle W_{UV} \rangle$ 
   and $\langle W_{UV}^{2} \rangle^{\frac{1}{2}}$), and 
   pulse height per plane ($\langle Q_{shw} \rangle$ and 
   $\langle Q_{shw}^{2} \rangle^{\frac{1}{2}}$), 
   are calculated. 
   These shower topology variables are required to satisfy:
   \begin{center}
   $\langle W_{UV} \rangle<5\,(4)$\,strips,\\ 
   $\langle W_{UV}^{2} \rangle^{\frac{1}{2}}<4\,(3)$\,strips,\\
   $\langle Q_{shw} \rangle<150\,(100)$\,PEs,\\
   $\langle Q_{shw}^{2} \rangle^{\frac{1}{2}}<150\,(100)$\,PEs,\\
   \end{center}
   for long (short) showers, respectively.
   Figure~\ref{fig_nue_topolgy_cut} 
   shows the predicted and observed distributions
   of the mean and RMS shower pulse height variables.
   An~analysis of the principal moments
   of the shower is also used to distinguish 
   the shower-like event topology of atmospheric neutrinos 
   from the track-like topology of cosmic-ray muons. 
   The moment of inertia tensor is constructed
   from the relative positions of the shower strips,
   weighted by their pulse height. The tensor 
   is diagonalized and selection criteria are placed 
   on the largest eigenvalue, $I_{UV}^{max}$,
   with long (short) showers required to satisfy
   $I_{UV}^{max}<0.15\,(0.05)$\,m$^{2}$, respectively.

 \item \textit{Removal of Selected Tracks: }
   After applying the above selection criteria, 
   it is found that 2\% of the resulting events 
   have been previously selected as contained-vertex tracks.
   Monte Carlo studies indicate that approximately half of
   the duplicate events are $\nu_{\mu}$ or $\overline{\nu}_{\mu}$ CC interactions.
   As a final step, these events are removed from 
   the contained-vertex shower sample. 
   
\end{enumerate}

The shower containment requirements select 701 events from the data,
compared with a total prediction of $727 \pm 101$ events 
in the absence of oscillations, and a prediction of $684 \pm 95$ events
for oscillations with $\Delta m^{2} = 2.32 \times 10^{-3} \mbox{\,eV}^{2}$ and $\mbox{sin}^{2} 2\theta = 1.0$.
The uncertainties are dominated by the 15\% uncertainty in the overall normalization, 
but also includes additional uncertainties 
of 20\% and 5\% in the NC and $\nu_{e}$+$\overline{\nu}_{e}$ CC 
components, respectively. 
The cosmic-ray muon prediction of $87 \pm 9$ events represents 
a background level of 12\%. Figure~\ref{fig_events_nue_energy} 
shows the predicted and observed energy distributions of the
selected events. Since the selected sample contains a small 
$\nu_{\mu}$+$\overline{\nu}_{\mu}$ CC component,
the distribution has only a weak dependence on the 
oscillation parameters.

\subsection{Summary of results from atmospheric neutrino event selection}
 
In total, 2072 candidate atmospheric neutrino events 
are selected from the data. 
For analysis, the events are grouped into:
905 contained-vertex muons, with vertex positions 
inside the fiducial volume (including both 
fully-contained and partially-contained muons); 
466 neutrino-induced rock-muons, 
with vertex positions outside the fiducial volume;
and the 701 contained-vertex showers. 
Table~\ref{table_event_selection}~gives the
predicted event rates for each of these samples.

\begin{table*}[!t]
 \begin{ruledtabular}
 \begin{tabular}{ccccccccc}
        & Data & \multicolumn{7}{c}{Prediction (no oscillations)} \\
        &      & Cosmic-ray $\mu$ & $\nu_{\mu}$+$\overline{\nu}_{\mu}$ CC & $\nu_{e}$+$\overline{\nu}_{e}$ CC & $\nu_{\tau}$+$\overline{\nu}_{\tau}$ CC & NC & Rock-$\mu$ & Total \\
 \hline 
 Contained-vertex muons  & $905$ & $34\pm3$ & $998\pm150$ & $35\pm6$ & $-$ & $25\pm6$ & $9\pm2$ & $1100\pm159$ \\
 Neutrino-induced rock-muons  & $466$ & $-$ & $26\pm4$ & $0\pm0$ & $-$ & $0\pm0$ & $544\pm136$ & $570\pm136$ \\
 Contained-vertex showers  & $701$ & $87\pm9$ & $157\pm24$ & $358\pm57$ & $-$ & $124\pm31$ & $1\pm0$ & $727\pm101$ \\
 \hline Total & 2072 & \multicolumn{7}{c}{ $2397\pm296 $} \\
 \hline
 \hline & Data & \multicolumn{7}{c}{Prediction ($\Delta m^{2} = 2.32 \times 10^{-3} \mbox{\,eV}^{2}, \mbox{sin}^{2} 2\theta = 1.0$)} \\
        &      & Cosmic-ray $\mu$ & $\nu_{\mu}$+$\overline{\nu}_{\mu}$ CC & $\nu_{e}$+$\overline{\nu}_{e}$ CC & $\nu_{\tau}$+$\overline{\nu}_{\tau}$ CC & NC & Rock-$\mu$ & Total \\
 \hline 
 Contained-vertex muons  & $905$ & $34\pm3$ & $689\pm103$ & $35\pm6$ & $3\pm1$ & $25\pm6$ & $6\pm1$ & $792\pm113$ \\
 Neutrino-induced rock-muons  & $466$ & $-$ & $14\pm2$ & $0\pm0$ & $0\pm0$ & $0\pm0$ & $433\pm108$ & $447\pm108$ \\
 Contained-vertex showers  & $701$ & $87\pm9$ & $110\pm16$ & $358\pm57$ & $5\pm2$ & $124\pm31$ & $0\pm0$ & $684\pm95$ \\
 \hline Total & 2072 & \multicolumn{7}{c}{ $1923\pm235 $} \\
 \end{tabular}
 \end{ruledtabular}
\caption{\label{table_event_selection}
  Summary of atmospheric neutrino selection, 
  separated into the different categories of selected and simulated event. 
  The Monte Carlo predictions are given separately for contained-vertex 
  atmospheric neutrinos and neutrino-induced rock-muons,
  with the contained-vertex predictions also separated by neutrino interaction type. 
  The cosmic muon background prediction is calculated directly from the data 
  by weighting vetoed events according to the measured shield efficiency.
  In the top table, the~predictions are calculated in the absence
  of neutrino oscillations; the~bottom table uses representative oscillation parameters
  of $\Delta m^{2} = 2.32 \times 10^{-3} \mbox{\,eV}^{2}$ and $\mbox{sin}^{2} 2\theta = 1.0$.
  The predictions and their uncertainties have been rounded to the nearest event.
  Note that many of the uncertainties are correlated and cancel in the ratios 
  and fits described in this paper.}
 \end{table*}

\subsection{Selection of high resolution event sample}

A sample of high resolution contained-vertex muons,
with well-measured muon propagation direction, 
is selected from the data. 
These high resolution events are required to 
satisfy minimum track length requirements of 10 planes and 1\,m.
They are also required to pass a timing requirement of $r_{L}-r_{H} < -0.66$\,ns. 
These selection criteria are found to correctly distinguish the 
track direction in 99\% of simulated atmospheric neutrinos.

A total of 631 high resolution contained-vertex muons are selected,
with the remaining 274 contained-vertex muons classified as 
low resolution. In the high resolution sample, 261 events
are classified as upward-going and 370 events are classified as downward-going. 
The measured up-down ratio 
is $R^{data}_{u/d}=0.71\pm0.06\,(\mbox{stat.})$, where the statistical 
error corresponds to the 68\% confidence interval calculated 
using Poisson statistics \cite{gehrels}. The predicted ratio, 
calculated from the simulation, in the absence 
of oscillations, is $R^{MC}_{u/d}=1.14 \pm 0.03\,(\mbox{syst.})$.
The 3\% systematic uncertainty combines the uncertainties 
in the event selection, and the atmospheric neutrino flux 
simulation. The double ratio between the observed and 
predicted up-down ratio is:

 \begin{equation*}
  R^{data}_{u/d}/R^{MC}_{u/d} = 0.62\pm0.05\,(\mbox{stat.})\pm0.02\,(\mbox{syst.}).
 \end{equation*}

\noindent
This ratio is in excess of 6 standard deviations from unity, 
indicating the presence of neutrino oscillations.

 \begin{table*}[!tb]
 \begin{ruledtabular}
 \begin{tabular}{cccccccc}
        & Data & \multicolumn{6}{c}{Prediction (no oscillations)} \\
        &      & $\nu_{\mu}$ CC & $\overline{\nu}_{\mu}$ CC & Rock-$\mu^{-}$ & Rock-$\mu^{+}$ & Other & Total \\
 \hline 
 Contained-vertex muons ($\mu^{-}$)  & $379$ & $425\pm64$ & $4\pm1$ & $4\pm1$ & $0\pm0$ & $13\pm1$ & $445\pm65$ \\
 Contained-vertex muons ($\mu^{+}$)  & $173$ & $12\pm2$ & $190\pm28$ & $0\pm0$ & $1\pm0$ & $15\pm2$ & $219\pm31$ \\
 \hline 
 Neutrino-induced rock-muons ($\mu^{-}$)  & $152$ & $16\pm2$ & $0\pm0$ & $215\pm54$ & $1\pm0$ & $0\pm0$ & $233\pm54$ \\
 Neutrino-induced rock-muons ($\mu^{+}$)  & $95$ & $0\pm0$ & $8\pm1$ & $3\pm1$ & $102\pm25$ & $0\pm0$ & $112\pm26$ \\
 \hline
 \hline & Data & \multicolumn{6}{c}{Prediction ($\Delta m^{2} = 2.32 \times 10^{-3} \mbox{\,eV}^{2}, \mbox{sin}^{2} 2\theta = 1.0$)} \\
        &      & $\nu_{\mu}$ CC & $\overline{\nu}_{\mu}$ CC & Rock-$\mu^{-}$ & Rock-$\mu^{+}$ & Other & Total \\
 \hline 
 Contained-vertex muons ($\mu^{-}$)  & $379$ & $294\pm44$ & $3\pm0$ & $2\pm1$ & $0\pm0$ & $15\pm2$ & $314\pm46$ \\
 Contained-vertex muons ($\mu^{+}$)  & $173$ & $9\pm1$ & $132\pm20$ & $0\pm0$ & $1\pm0$ & $16\pm2$ & $158\pm22$ \\
 \hline 
 Neutrino-induced rock-muons ($\mu^{-}$)   & $152$ & $9\pm1$ & $0\pm0$ & $151\pm38$ & $1\pm0$ & $0\pm0$ & $161\pm38$ \\
 Neutrino-induced rock-muons ($\mu^{+}$)  & $95$ & $0\pm0$ & $4\pm1$ & $2\pm0$ & $68\pm17$ & $0\pm0$ & $74\pm18$ \\
 \end{tabular}
 \end{ruledtabular}
\caption{\label{table_charge_separation}
  Results from the separation of contained-vertex and 
  neutrino-induced rock-muons by reconstructed charge sign, 
  into selected samples of neutrinos and antineutrinos.
  The predictions from each simulated sample are given separately 
  for true neutrinos and antineutrinos; the~column labelled 
  `Other' is the sum of the cosmic-ray muon,
  $\nu_{e}$+$\overline{\nu}_{e}$, $\nu_{\tau}$+$\overline{\nu}_{\tau}$ and NC backgrounds.
  All the predictions and their uncertainties have been rounded to the nearest event.
  Note that many of the uncertainties are correlated and cancel in the 
  ratios and fits described in this paper.}
\end{table*}

\section{\label{ChargeRatio}Separation of neutrinos and antineutrinos}

The high resolution contained-vertex muon sample
and neutrino-induced rock-muon sample
are separated into candidate neutrinos and antineutrinos 
based on the reconstructed muon charge sign. 
The Kalman filter returns a best fit value of $q/p$, 
and its uncertainty, $\sigma_{q/p}$, 
where $q$ is the muon charge sign and $p$ is the muon momentum.
The selected events are classified as neutrinos if $q<0$, 
and as antineutrinos if $q>0$.

\begin{figure*}[!p]
   \includegraphics[width=\textwidth]{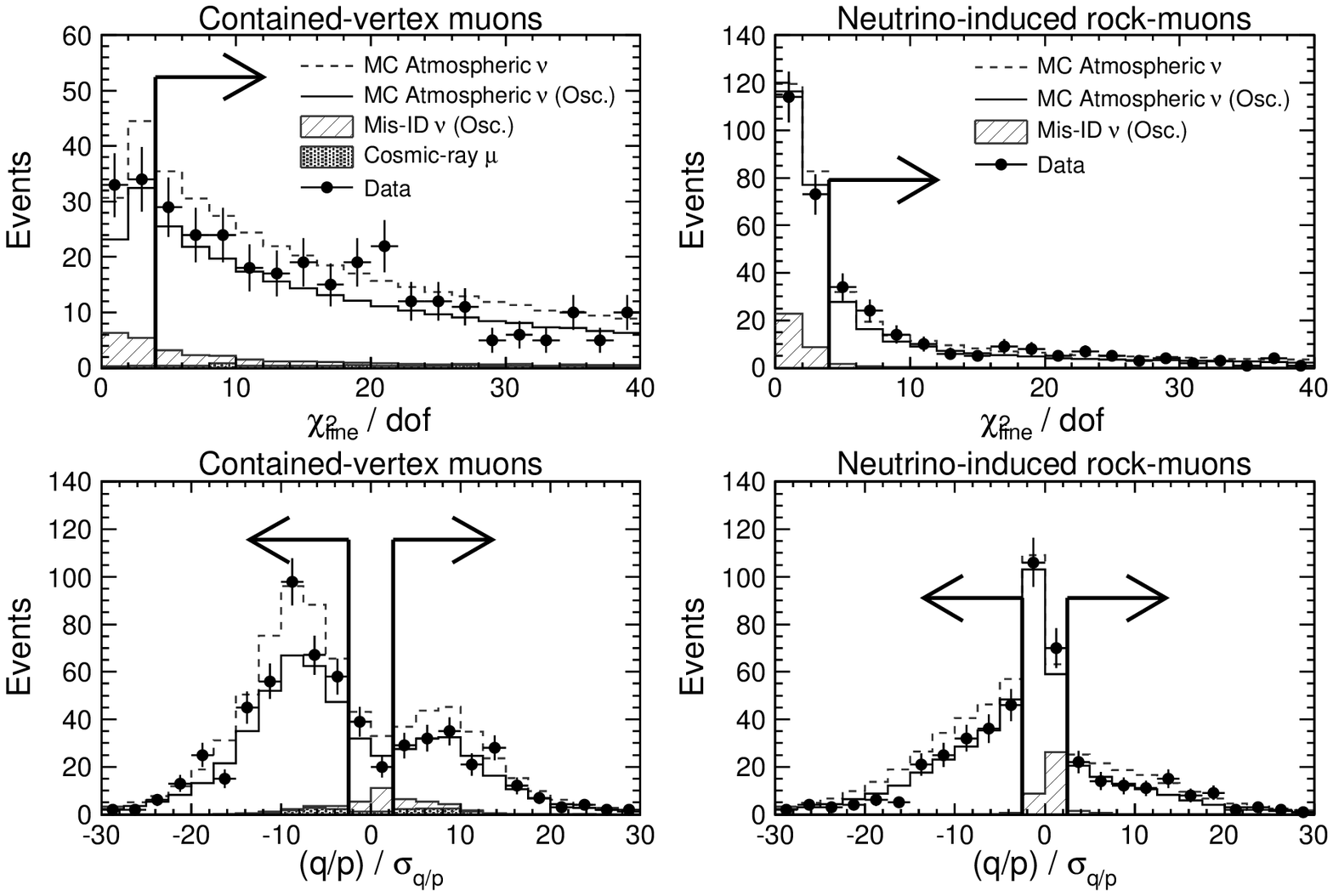}
   \caption{\label{fig_charge_separation}
     Distributions of track fit uncertainty, $(q/p)/\sigma_{q/p}$
     and track straightness variable, $\chi^{2}_{line}/dof$, 
     used to select events with well-measured muon charge sign.
     The distributions are plotted for contained-vertex muons (top panels),
     and neutrino-induced rock-muons (bottom panels).
     In each plot, the dashed line indicates the total prediction in the absence
     of oscillations; the solid line shows the prediction for oscillations with
     $\Delta m^{2} = 2.32 \times 10^{-3} \mbox{\,eV}^{2}$ 
     and $\mbox{sin}^{2} 2\theta = 1.0$; the shaded histogram shows
     the cosmic-ray muon background; 
     and the points show the observed data. 
     In addition, the hatched histograms show the component
     with mis-identified charge sign.
     The arrows indicate the selections used to identify
     events with well-measured charge sign.} 
  \vspace{0.25cm}
   \includegraphics[width=\textwidth]{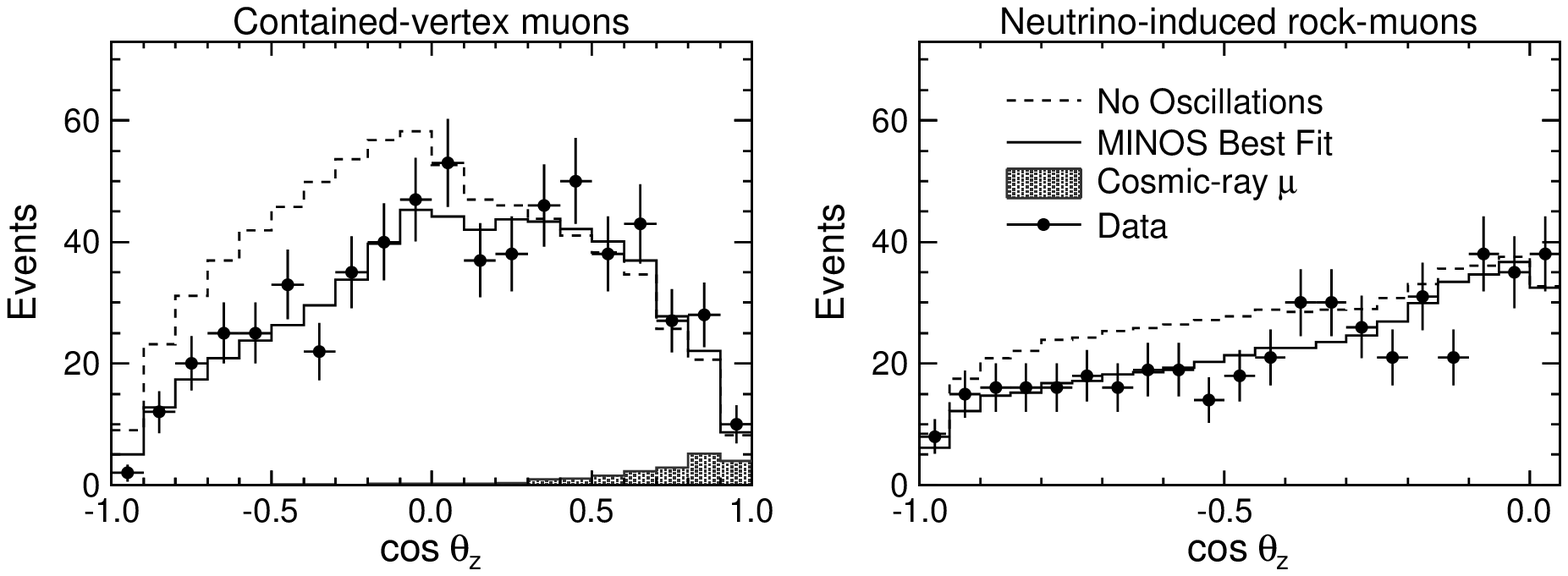} 
   \caption{\label{fig_data_zenith_angle}
     Distributions of reconstructed zenith angle,
     for contained-vertex muons (left),
     and neutrino-induced rock-muons (right). In~each plot, the dashed line 
     gives the nominal prediction for the case of no oscillations; 
     the shaded histogram shows the cosmic-ray muon background;
     and the points with errors show the observed data.
     The solid line shows the best fit to the data,
     which combines the best fit oscillation and systematic parameters.} 
 \end{figure*}

Two criteria are used to select events with significant 
track curvature and therefore well-measured charge sign.
First, a requirement is placed on the relative size of the track fit uncertainty,
$|q/p|/\sigma_{q/p}$, which indicates the significance 
of the track curvature \cite{minoscontainedvertex}. 
Events are required to satisfy $|q/p|/\sigma_{q/p} > 2.5$. 
The reconstructed track is then used to calculate a variable 
measuring the straightness of the track, with the aim 
of excluding tracks which do not have significant curvature.
A straight line is drawn between the reconstructed start 
and end point of the track,
and a chi-squared variable, $\chi^{2}_{line}/dof$, 
is calculated from the deviations of the track strips 
from this line \cite{minosupwardmuon}. Events are required 
to satisfy $\chi^{2}_{line}/dof > 4.0$. 
Figure~\ref{fig_charge_separation} shows the observed and 
predicted distributions of these two selection variables 
for contained-vertex muons and neutrino-induced rock-muons.

The charge selection criteria are found to correctly identify 
the muon charge in 97\% of simulated contained-vertex interactions,
and 99\% of simulated neutrino-induced rock-muons. 
The selection efficiencies are 87\% and 59\%, respectively,
calculated as a fraction of the number of selected events 
with well-measured direction. 
The lower efficiency for neutrino-induced rock-muons 
reflects their higher average momentum, resulting in 
more events with ambiguous track curvature.

Table~\ref{table_charge_separation} gives the 
predicted and observed numbers of neutrinos and 
antineutrinos for each category of event. 
For contained-vertex muons, the charge-separation procedure
returns 379 neutrinos and 173 antineutrinos, 
giving a measured charge ratio of 
$R^{data}_{\overline{\nu}/\nu}=0.46^{+0.05}_{-0.04}\,(\mbox{stat.})$.
The predicted value of the charge ratio is calculated 
from the simulation to be
$R^{MC}_{\overline{\nu}/\nu}=0.49 \pm 0.05\,(\mbox{syst.})$.
The prediction is almost entirely independent 
of any input oscillations, provided that equal 
parameters are used for neutrinos and antineutrinos.
The overall systematic uncertainty of 10\% is obtained by 
combining uncertainties of 8.5\% in the ratio of the
neutrino and antineutrino interaction cross-sections, 
4\% in the flux ratio of neutrinos and antineutrinos, 
and 3\% in the purity of the charge separation. 
The double ratio between the observed and predicted 
charge ratios is calculated to be:
  $R^{data}_{\overline{\nu}/\nu}/R^{MC}_{\overline{\nu}/\nu} 
    = 0.93 \pm 0.09 \,(\mbox{stat.}) \pm 0.09\,(\mbox{syst.})$.

For neutrino-induced rock-muons, 152 neutrinos and 95 antineutrinos
are selected, giving a measured 
charge ratio of $R^{data}_{\overline{\nu}/\nu}=0.63^{+0.09}_{-0.08}\,(\mbox{stat.})$.
The predicted value of the charge ratio is calculated 
from the simulation to be 
$R^{MC}_{\overline{\nu}/\nu}=0.48 \pm 0.06\,(\mbox{syst.})$.
The overall systematic uncertainty of 12.5\% is obtained by 
combining uncertainties of 10\% in the flux ratio, 
4\% in the cross-section ratio, and 6\% in the 
charge-separation purity. The double ratio between 
the observed and predicted charge ratio is:
  $R^{data}_{\overline{\nu}/\nu}/R^{MC}_{\overline{\nu}/\nu} 
   = 1.29^{+0.19}_{-0.17}\,(\mbox{stat.}) \pm 0.16\,(\mbox{syst.})$.

The charge-separated samples of contained-vertex muons and neutrino-induced rock-muons 
are combined to give an overall double ratio of:

 \begin{equation*}
  R^{data}_{\overline{\nu}/\nu}/R^{MC}_{\overline{\nu}/\nu} 
   = 1.03 \pm 0.08\,(\mbox{stat.}) \pm 0.08\,(\mbox{syst.}).
 \end{equation*}

\noindent
This result is consistent with unity.

 \begin{figure*}[!p]
   \includegraphics[width=\textwidth]{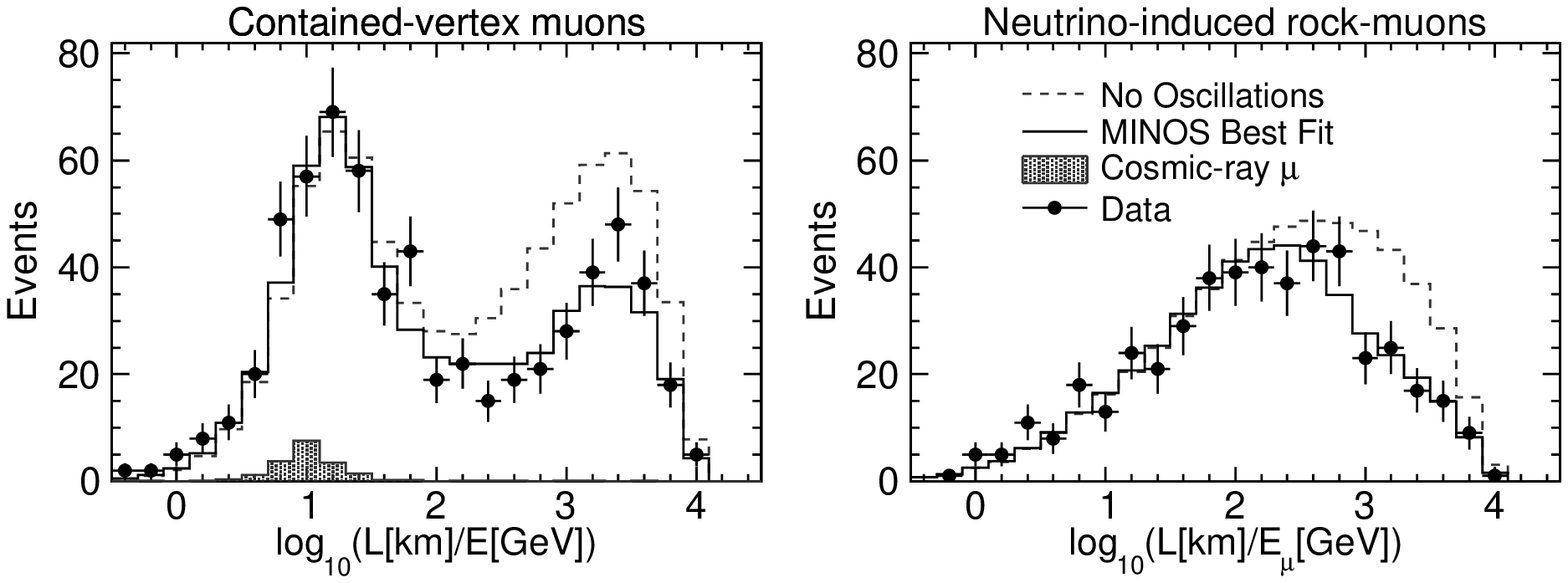} 
   \caption{\label{fig_data_oscfit_qcomb}
     Distributions of reconstructed $\mbox{log}_{10}(L/E)$,
     for contained-vertex muons (left),
     and neutrino-induced rock-muons (right). In~each plot, the dashed line 
     gives the nominal prediction for the case of no oscillations; 
     the shaded histogram shows the cosmic-ray muon background;
     and the points with errors show the observed data.
     The solid line shows the best fit to the data,
     which combines the best fit oscillation and systematic parameters.} 
   \vspace{0.25cm}
   \includegraphics[width=\textwidth]{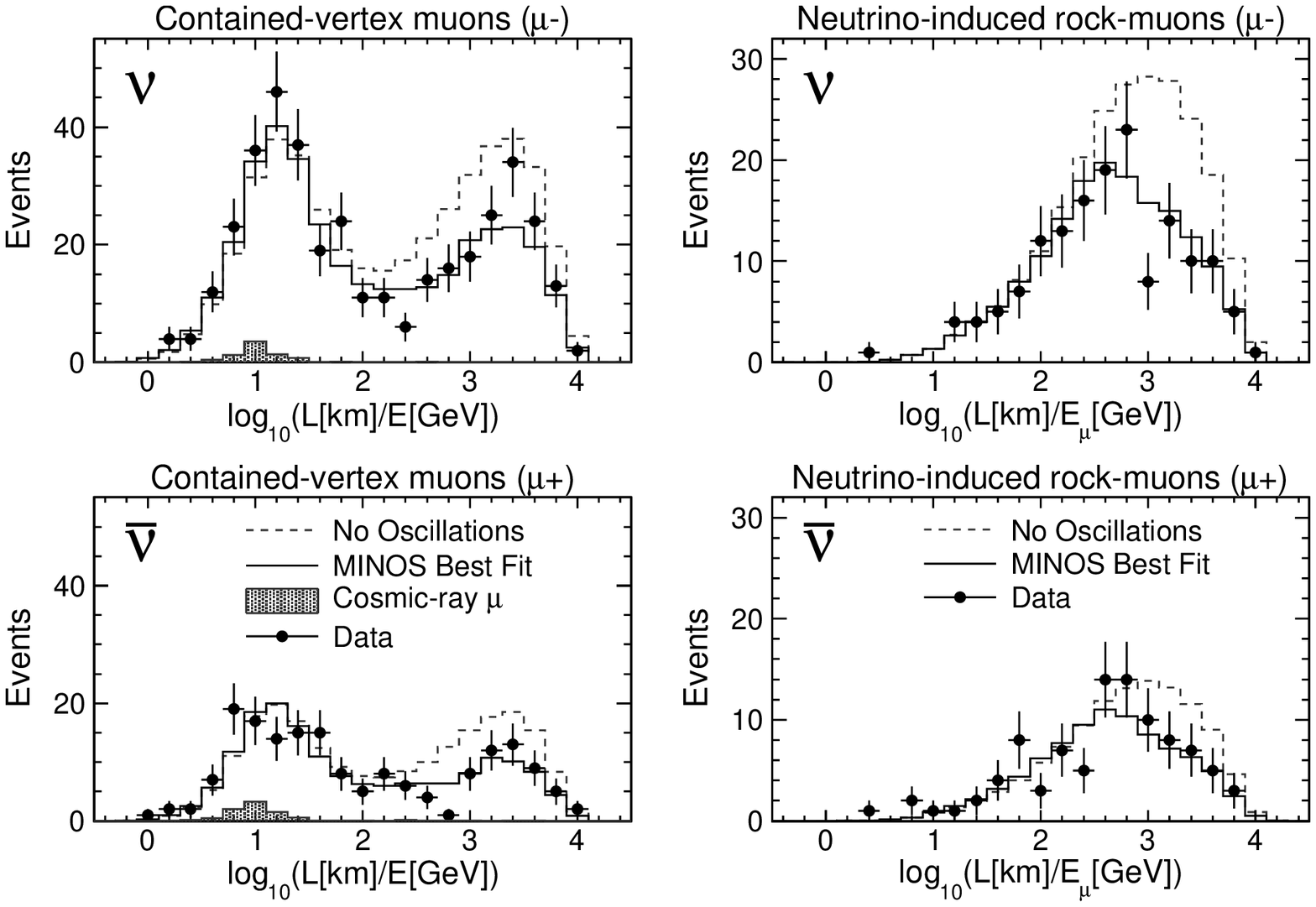}
   \caption{\label{fig_data_oscfit_qsep}
     Distributions of reconstructed $\mbox{log}_{10}(L/E)$, 
     plotted for selected neutrinos and antineutrinos in the 
     contained-vertex muon sample (top panels) and neutrino-induced
     rock-muon sample (bottom panels).
     In each case, the dashed line gives the the nominal prediction
     in the absence of oscillations; the shaded histogram shows the 
     cosmic-ray muon background; and the points with errors show 
     the observed data. The solid line indicates the best fit to the data,
     which combines the best fit oscillation and systematic parameters.} 
 \end{figure*}

\section{\label{OscillationAnalysis}Oscillation analysis}

Two oscillation fits are applied to the selected data.
The first is a two-parameter fit, which outputs equal oscillation parameters 
for neutrinos and antineutrinos; 
the second is a four-parameter fit, which outputs separate oscillation parameters 
for neutrinos and antineutrinos.
Each fit is applied to the reconstructed
$L/E$ distributions of selected neutrinos and antineutrinos.
The neutrino propagation length, $L$, is determined
from the reconstructed zenith angle of the muon track.
For contained-vertex muons, the parent neutrino energy, $E$, 
is found by summing the reconstructed muon energy and visible shower energy. 
For neutrino-induced rock-muons, only the muon energy is used, 
due to the increased uncertainties in the 
shower simulation and calibration at the edge of the detector,
or because the interaction occurs outside the detector 
and hence the vertex is not visible. 
Figure~\ref{fig_data_zenith_angle}~shows the predicted and observed 
reconstructed zenith angle distributions for contained-vertex muons 
and neutrino-induced rock-muons.
Figures~\ref{fig_data_oscfit_qcomb}~and~\ref{fig_data_oscfit_qsep}
show the predicted and observed $L/E$ distributions for these two samples,
before and after the separation of events into neutrinos and antineutrinos.
For each of these figures, the observed data
are compared with the best fit neutrino and antineutrino
oscillation parameters, which are given in Section~\ref{TestingCPT}.

\subsection{Separation of events by L/E resolution}

The intrinsic $L/E$ resolution of contained-vertex events,
and therefore the degree to which they contribute to the 
overall sensitivity to the oscillation parameters,
varies significantly across the high resolution sample. 
The resolution in the propagation distance
depends on the reconstructed 
energy and zenith angle, and is worse at low energies
where the average angle between the neutrino and muon is large, 
and also around the horizon where the propagation distance
varies rapidly as a function of zenith angle. 
The resolution in the neutrino energy is worse for events 
where the muon momentum is determined from curvature rather 
than range, and also for events with high $y$ values, 
since the shower energy resolution is generally 
poorer than the muon momentum resolution. 

The sensitivity to oscillations can be improved by
incorporating information on $L/E$ resolution
into the oscillation fit.
For this analysis, a Bayesian technique is used
to estimate the $L/E$ resolution of selected
contained-vertex muons on an event-by-event basis.
For each event, a probability distribution function (PDF)
in $\mbox{log}_{10}(L/E)$ is calculated by combining the
measured muon momentum, muon direction and shower energy
of the event
with information from the Monte Carlo simulation 
describing the atmospheric neutrino spectrum,
interaction kinematics and detector resolution.
The simulation is used to construct PDFs relating the
measured muon momentum and shower energy of selected
$\nu_{\mu}$ and $\overline{\nu}_{\mu}$ events to their 
corresponding true distributions.
The simulation also provides PDFs 
in the kinematic variables $W^{2}$ and $y$ for 
$\nu_{\mu}$ and $\overline{\nu}_{\mu}$ CC interactions, 
binned as a function of neutrino energy,
which enable the distributions of muon momentum and
shower energy to be mapped onto a distribution of
neutrino energy, and the muon direction to be mapped
onto a distribution of neutrino zenith angle.
The overall PDF in $\mbox{log}_{10}(L/E)$ is then obtained
by taking a convolution of these neutrino distributions
and the RMS of this PDF, $\sigma_{log(L/E)}$, 
gives the $L/E$ resolution.
A full description of the technique 
is given in~\cite{bayesian}.

\begin{figure}[!p]
   \includegraphics[width=\columnwidth]{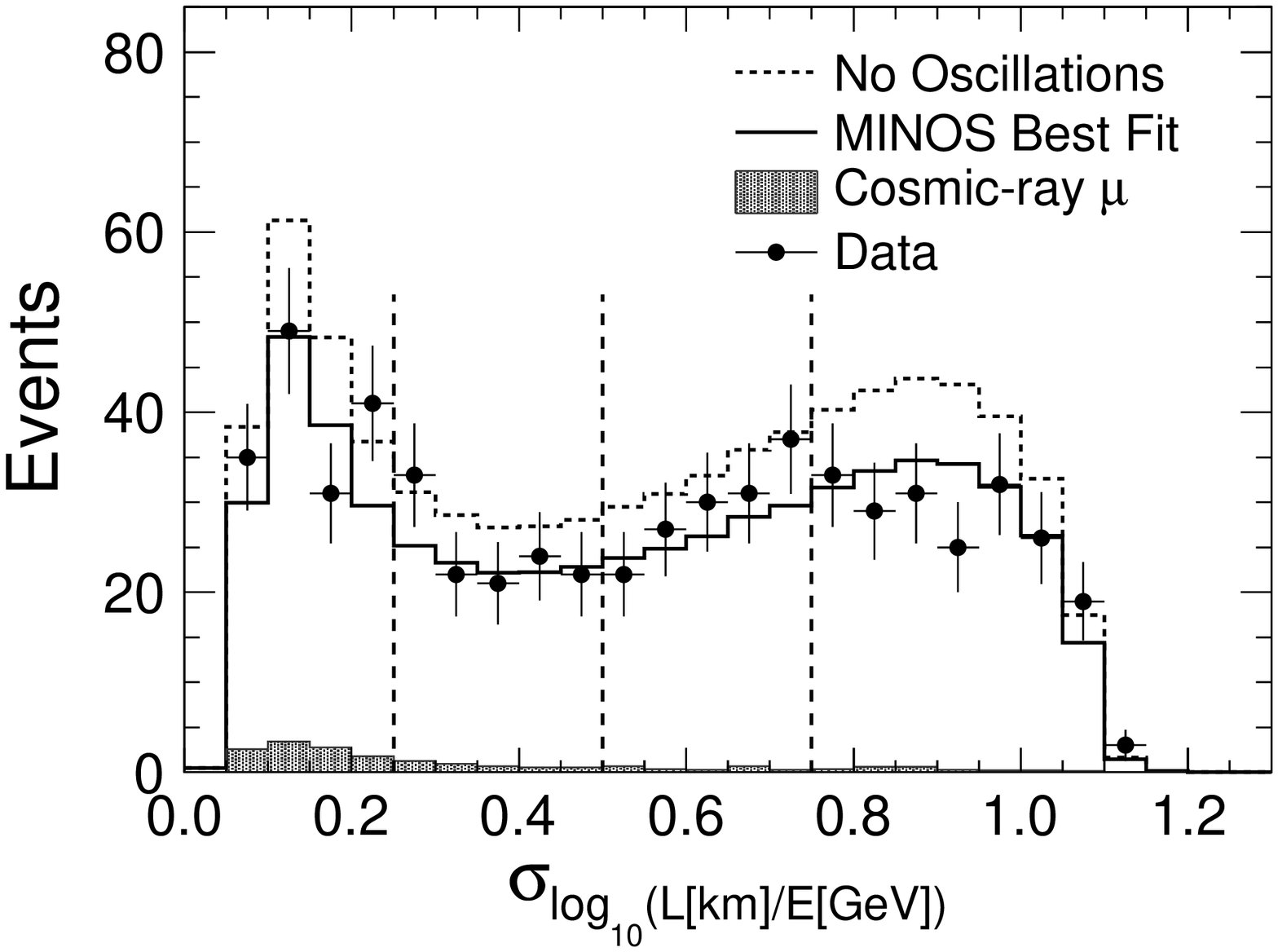}
   \caption{\label{fig_bayes_resolution}
     Distributions showing the calculated $L/E$ resolution
     for the high resolution sample of contained-vertex muon,
     which have well-measured propagation direction. 
     The dashed histogram indicates the nominal prediction in the absence of oscillations; 
     the solid histogram indicates the prediction for the
     best fit oscillation parameters presented in this paper;
     the shaded histogram shows the cosmic-ray muon background;
     and the points with error bars represent the observed data.
     The vertical dashed lines correspond to the partitions used
     to divide the selected events into four bins of $L/E$ resolution.} 
   \vspace{0.66cm}
   \includegraphics[width=\columnwidth]{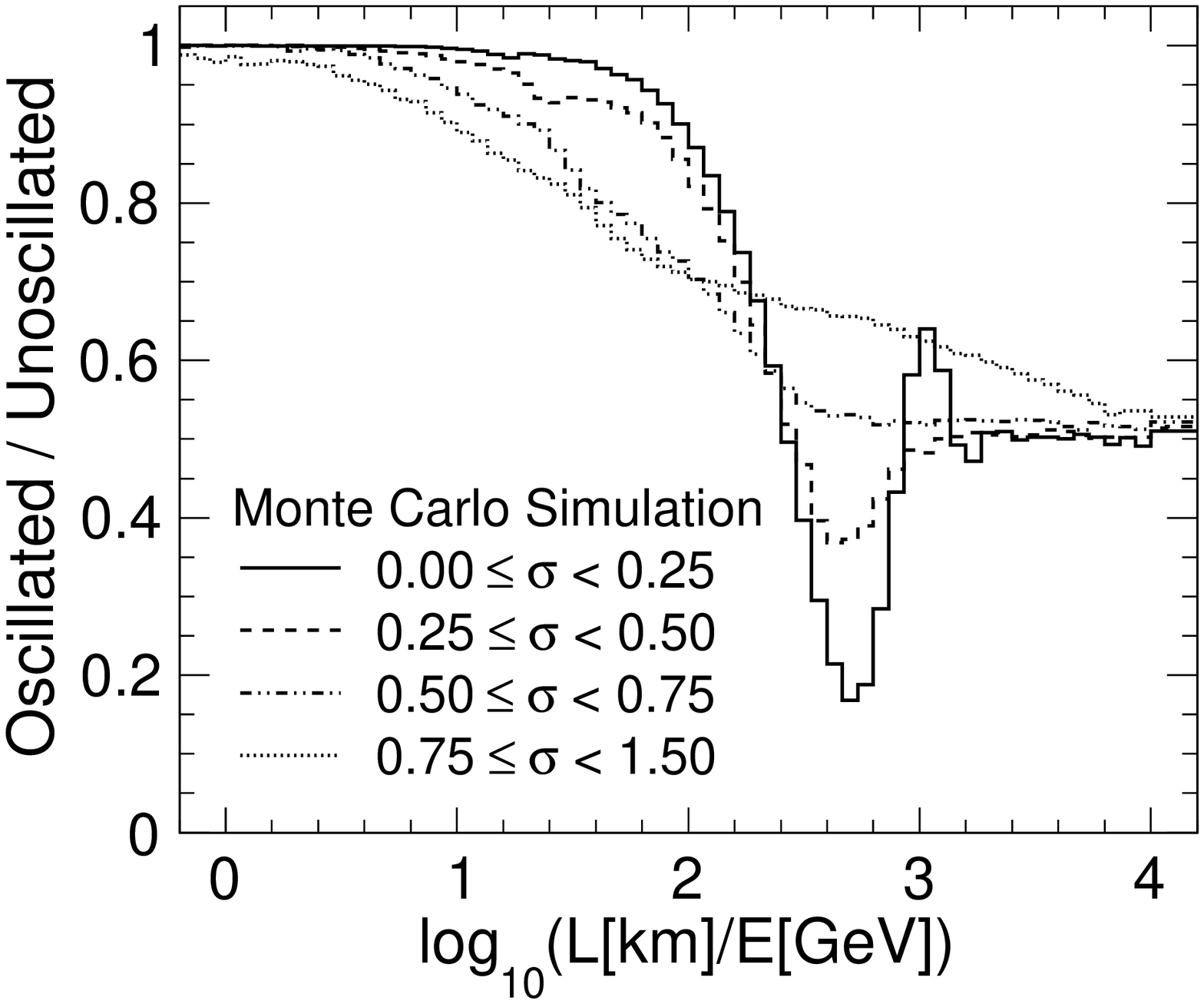}
   \caption{\label{fig_bayes_logle_mc_ratio}
     Ratios of the predicted $\mbox{log}_{10}(L/E)$ distributions 
     with oscillations to those without oscillations in the four bins 
     of $L/E$ resolution. The predictions with oscillations are generated 
     using input parameters of $\Delta m^{2} = 2.32 \times 10^{-3} \mbox{\,eV}^{2}$ 
     and $\mbox{sin}^{2} 2\theta = 1.0$. The oscillations are most
     sharply defined in the bin of highest resolution.
     Here, a clear oscillation dip can be seen at $\mbox{log}_{10}(L/E)\approx2.7$,
     corresponding to the peak oscillation probability.     
     The ratio then rises to a maximum at $\mbox{log}_{10}(L/E)\approx3$,
     and a second dip is visible before
     the ratio averages to $1-\frac{1}{2}\mbox{sin}^{2} 2\theta = 0.5$,
     as the frequency of oscillations becomes rapid.} 
 \end{figure}

Figure~\ref{fig_bayes_resolution} shows the predicted
and observed $\sigma_{log(L/E)}$ distributions
for high resolution contained-vertex muon neutrinos.
The shape of the predicted distribution is almost independent 
of the input oscillation parameters. The observed spread 
of $\sigma_{log(L/E)}$ values is substantial, 
and corresponds to $25$\% of the spread in 
$\mbox{log}_{10}(L/E)$. 
Therefore, 
a significant gain in sensitivity is expected 
by separating events into bins of $L/E$ resolution.

For the oscillation analysis, the selected 
contained-vertex muon neutrinos
are divided into the following four bins of 
$L/E$ resolution:

\begin{center}
$0.00\leq\sigma_{log(L/E)}<0.25$,\\
$0.25\leq\sigma_{log(L/E)}<0.50$,\\ 
$0.50\leq\sigma_{log(L/E)}<0.75$,\\
$0.75\leq\sigma_{log(L/E)}<1.50$.
\end{center}

\noindent
Figure~\ref{fig_bayes_logle_mc_ratio} shows the ratio 
of the predicted $L/E$ distributions with oscillations 
to those without oscillations in each bin of resolution.
The oscillations are most sharply 
resolved in the bin with the best $L/E$ resolution. 
Here, the ratio initially falls with $L/E$, 
reaching a minimum at the peak oscillation probability.
The ratio subsequently rises to a maximum, 
and a second oscillation dip is visible
before the ratio averages to $1-\frac{1}{2}\mbox{sin}^{2}2\theta$
as the frequency of oscillations becomes rapid.

Selected neutrino-induced rock-muons are separated into low momentum 
($P_{\mu}\leq10$\,GeV) and high momentum ($P_{\mu}>10$\,GeV) 
samples. This separation roughly distinguishes those 
muons whose parent neutrinos have a relatively large 
oscillation probability from those with a lower probability. 

\begin{figure*}[!t]    
   \includegraphics[width=\textwidth]{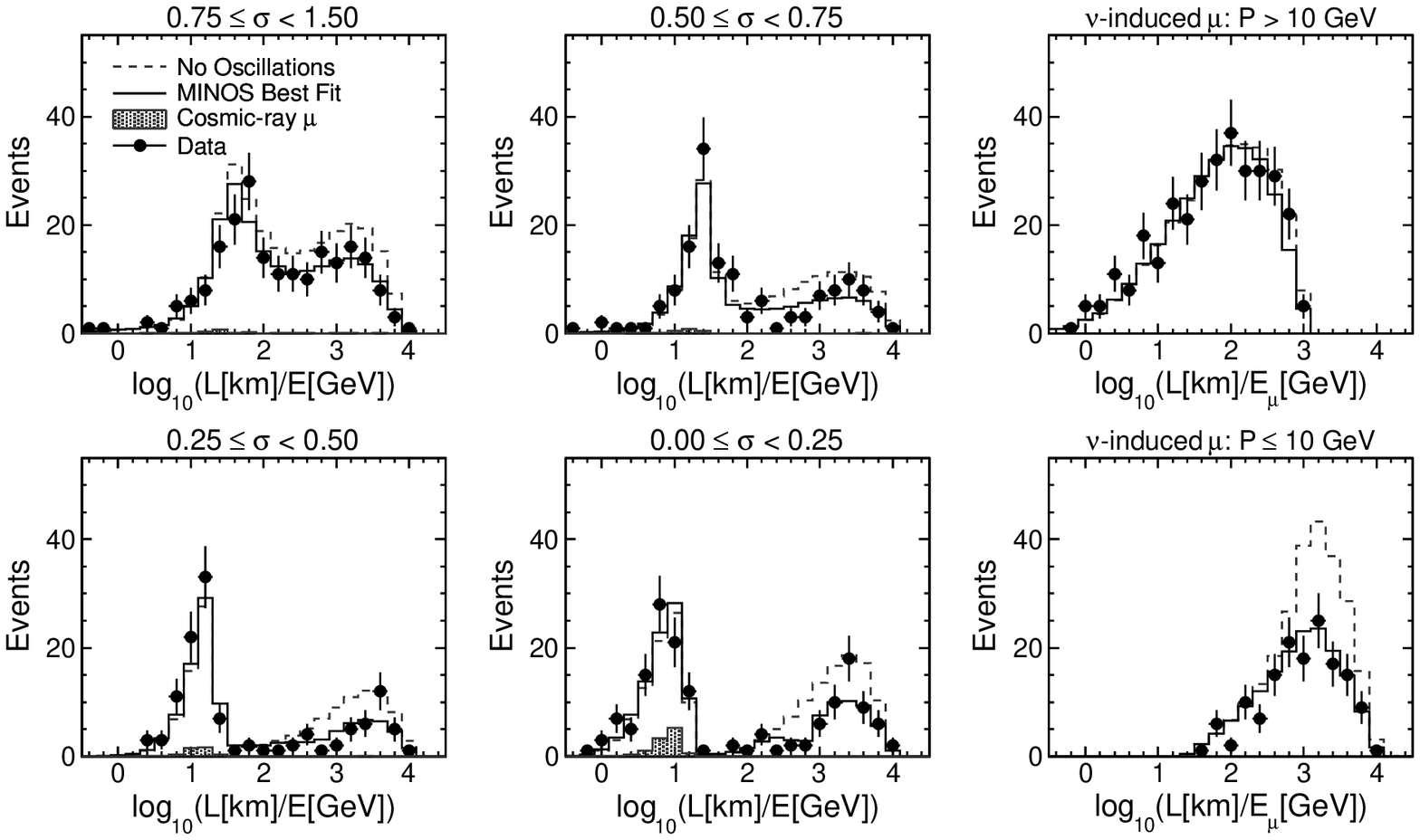}
   \caption{\label{fig_data_oscfit_all_fcpc}
     Distributions of $\mbox{log}_{10}(L/E)$ observed in
     each bin of $L/E$ resolution for contained-vertex muons, 
     and in each bin of muon momentum for neutrino-induced rock-muons.
     For each of the panels,
     the dashed line gives the the nominal prediction in the 
     absence of oscillations; the shaded histogram shows the 
     cosmic-ray muon background; and the points represent the data. 
     The solid line indicates the best fit to the data,
     combining the best fit oscillation and systematic parameters.} 
 \end{figure*}

Figure~\ref{fig_data_oscfit_all_fcpc} shows the predicted and 
observed $L/E$ distributions, separated into bins of $L/E$ resolution 
for contained-vertex muons, and into bins of muon momentum 
for neutrino-induced rock-muons. The predicted distributions are
calculated for the case of no oscillations, and for the
best fit neutrino and antineutrino oscillation parameters.

\subsection{Oscillation fit}

An oscillation fit is applied to the data assuming 
two-flavor $\nu_{\mu} \rightarrow \nu_{\tau}$ vacuum 
oscillations. In this approximation, the oscillation 
probability is given by:

 \begin{equation*}
  P \left( \nu_{\mu} \rightarrow \nu_{\tau} \right) = \mbox{sin}^{2} 2\theta \mbox{ } \mbox{sin}^{2} \left( \frac{ \Delta m^{2} \mbox{ } \mbox{L} }{ 4 \mbox{ } \mbox{E} } \right),
 \end{equation*}

\noindent
where $\Delta m^{2}$ and sin$^{2}2\theta$ are
the two-flavor oscillation parameters,
$L$ is the neutrino propagation distance, 
and $E$ is the neutrino energy. 

For upward-going atmospheric neutrinos with energies
in the $2-10$\,GeV region, 
an asymmetry between muon neutrinos and antineutrinos
is predicted, arising from the resonant enhancement of
three-flavor oscillations by matter effects~\cite{matter1}.
The sign of the asymmetry depends on the sign of the 
neutrino squared-mass difference, and is therefore 
sensitive to the mass ordering of neutrinos~\cite{matter2}.
These effects have only a small influence on the predicted 
$L/E$ distributions of neutrinos and antineutrinos
and so are not considered in this analysis.

The high and low resolution contained-vertex muons, 
neutrino-induced rock-muons, and contained-vertex showers 
are each included as separate samples in the oscillation fit.
For the low resolution contained-vertex muons and contained-vertex shower samples,
the events are fitted in single bins of normalization.
The high resolution contained-vertex muon sample is divided into
two bins of direction, $r$\,=\,($u$,\,$d$), corresponding to upward-going ($u$)
and downward-going ($d$) muons;
three bins of charge sign, $s$\,=\,($\nu$,\,$\overline{\nu}$,\,$X$),
corresponding to neutrinos ($\nu$), antineutrinos ($\overline{\nu}$)
and events with ambiguous charge sign ($X$);
and four bins of $L/E$ resolution.
The neutrino-induced rock-muon sample is divided into three bins of charge sign, 
and two bins of muon momentum.
Overall, there are $24$ high resolution contained-vertex muon 
and $6$ neutrino-induced rock-muon distributions.
Each distribution is binned in $\mbox{log}_{10}(L/E)$, 
using 25 bins in the range $[-0.5,+4.5]$. 
This gives a total of 750 high resolution bins,
in addition to the 2 low resolution bins.

A maximum likelihood fit to the data is performed using
the following negative log-likelihood function:

\begin{equation*}
 \begin{split}
    -\ln {\cal{L}} &= \sum_{l} \mu - n \ln \mu + \sum_{h} \sum_{r,s} \mu - n \ln \mu\\ 
    &- \sum_{h} \sum_{r,s} \sum_{i,k} n_{ik} \ln \left(f_{ik}\right)\\ 
    &+ \sum_{j} \frac{\alpha_j^2}{2\sigma_{\alpha_j}^2}
  \end{split}
\end{equation*}

\noindent
This log-likelihood function is divided into the following terms:

\begin{enumerate}

 \item \textit{Normalization:} 
   The sums $\sum \mu - n \mbox{ ln\,} \mu$ represent the 
   Poisson probability for observing a total of $n$ events 
   with a prediction of $\mu$ events. The first sum,
   denoted $l$, is taken over the contained-vertex shower 
   and low resolution contained-vertex muon 
   samples, which are fitted in single bins; the second sum, denoted $h$, 
   is taken over the neutrino-induced rock-muon and high resolution contained-vertex muon samples, 
   which are separated by muon direction $r$\,=\,($u$,\,$d$) 
   and charge sign $s$\,=\,($\nu$,\,$\overline{\nu}$,\,$X$).

 \item \textit{Shape Term:} 
   The shape of the $\mbox{log}_{10}(L/E)$ distribution is 
   incorporated into the oscillation fit for 
   the neutrino-induced rock-muon and high resolution contained-vertex muon samples.
   The terms $\sum_{i,k} n_{ik} \ln \left(f_{ik} \right)$ 
   represent the likelihood functions for each of the $\mbox{log}_{10}(L/E)$
   distributions included in the fit. The $i$-sum is taken 
   over each resolution bin for the contained-vertex muons, 
   and each momentum bin for neutrino-induced rock-muons; 
   the $k$-sum is taken over each of
   the 25 bins in the $\mbox{log}_{10}(L/E)$ distribution.
   Within the sum, $n_{ik}$ is the observed number of events 
   and $f_{ik}$ is the relative predicted probability in
   the $i^{th}$ and $k^{th}$ bins.

 \item \textit{Systematic Uncertainties:} 
   Systematic effects are incorporated as nuisance parameters, 
   where the shift $\alpha_{j}$ is the deviation of the $j^{th}$ 
   systematic parameter from its nominal value. 
   A~penalty term, $\alpha_j^2/2\sigma_{\alpha_j}^2$, 
   is added to the likelihood, where the error $\sigma_{\alpha_j}$ 
   represents the estimated uncertainty in the $j^{th}$ 
   systematic parameter.

\end{enumerate}

A total of 12 systematic uncertainties
are incorporated into the fit as nuisance parameters,
as listed in Table~\ref{table_bestfits}.
For contained-vertex neutrino interactions, 
a 15\% uncertainty is applied to the normalization
of the event sample.
The following additional uncertainties are applied to this sample:
a 3\% uncertainty on the up-down ratio; 
a 5\% uncertainty on the ($\nu_{\mu}$+$\overline{\nu}_{\mu}$)/($\nu_{e}$+$\overline{\nu}_{e}$) ratio;
a 10\% uncertainty on the $\overline{\nu}_{\mu}$/$\nu_{\mu}$ ratio;
and a 20\% uncertainty on the ratio of NC to CC interactions.
For neutrino-induced rock-muons, a 25\% uncertainty
is applied on the normalization of the event sample.
An additional 12.5\% uncertainty is applied to the
$\overline{\nu}_{\mu}$/$\nu_{\mu}$ ratio in this sample.

To account for the uncertainty in the shape of the
atmospheric neutrino energy spectrum, the number of contained-vertex 
events is allowed to scale as a function of neutrino energy.
The form of the scaling function is chosen to cover the 
variations in the spectrum generated by changing the  
flux model and by reweighting the cross-section model 
according to its given uncertainties.
Above 3\,GeV, where the prediction from the simulation approximately 
follows a power function, 
events are scaled by $f(E_{\nu})=1+\alpha \mbox{ ln}(E_{\nu}/3)$.
Below 3\,GeV, this is connected smoothly to a linear function,
$f(E_{\nu})=1+\alpha \mbox{ }(E_{\nu}-3)$. 
The scaling function is applied separately to neutrinos and antineutrinos.
In each case, the spectrum parameter, $\alpha$, 
is normally distributed with a standard deviation of $6\%$. 
Finally, to account for the systematic uncertainties on the track and 
shower energy scale, a 3\% uncertainty is included on the 
muon momentum from range, 5\% on the momentum from curvature,
and 15\% on the shower energy scale. 
Of all the systematic uncertainties incorporated into the fit, 
only the two normalization parameters are found to have a significant impact 
on the resulting confidence limits.

 \begin{table*}[!b]
 \begin{ruledtabular}
 \begin{tabular}{cccc}
          Parameter                                           & Uncertainty       & Best Fit (2 Osc. Param.) & Best Fit (4 Osc. Param.) \\
   \hline $|\Delta m^{2}|/\mbox{eV}^{2}$                      & $$                & $1.9\times10^{-3}$ & $2.2\times10^{-3}$ \\
          $|\Delta \overline{m}^{2}|/\mbox{eV}^{2}$           & $$                & $1.9\times10^{-3}$ & $1.6\times10^{-3}$ \\
          $\mbox{sin}^{2} 2\theta$                            & $$                & $0.99$             & $0.99$             \\
          $\mbox{sin}^{2} 2\overline{\theta}$                 & $$                & $0.99$             & $1.00$             \\
   \hline normalization (contained-vertex $\nu$)              & $\sigma=15\,\%$   & $+0.6\,\sigma$     & $+0.7\,\sigma$ \\
          normalization ($\nu$-induced rock-$\mu$)            & $\sigma=25\,\%$   & $+0.1\,\sigma$     & $+0.1\,\sigma$ \\
          $up$/$down$ ratio (contained-vertex $\nu$)          & $\sigma=3\,\%$    & $-0.1\,\sigma$     & $-0.1\,\sigma$ \\
          $\nu_{e}$/$\nu_{\mu}$ ratio (contained-vertex $\nu$)                  & $\sigma=5\,\%$    & $-0.5\,\sigma$     & $-0.5\,\sigma$ \\
          $\overline{\nu}_{\mu}$/$\nu_{\mu}$ ratio (contained-vertex $\nu$) & $\sigma=10\,\%$   & $-0.5\,\sigma$     & $-0.6\,\sigma$ \\
          $\overline{\nu}_{\mu}$/$\nu_{\mu}$ ratio ($\nu$-induced rock-$\mu$) & $\sigma=12.5\,\%$ & $+1.1\,\sigma$     & $+0.9\,\sigma$ \\
          $NC/CC$ ratio (contained-vertex $\nu$)              & $\sigma=20\,\%$   & $+0.6\,\sigma$     & $+0.6\,\sigma$ \\
          $\nu$ spectrum parameter                            & $\sigma=6\,\%$    & $-0.4\,\sigma$     & $-0.4\,\sigma$ \\
          $\overline{\nu}$ spectrum parameter                 & $\sigma=6\,\%$    & $+0.3\,\sigma$     & $+0.3\,\sigma$ \\
          $\mu$ momentum (range)                              & $\sigma=3\,\%$    & $-0.3\,\sigma$     & $-0.3\,\sigma$ \\
          $\mu$ momentum (curvature)                          & $\sigma=5\,\%$    & $+0.3\,\sigma$     & $+0.3\,\sigma$ \\
          shower energy                                       & $\sigma=15\,\%$   & $+0.4\,\sigma$     & $+0.4\,\sigma$ \\
 \end{tabular}
 \end{ruledtabular}
 \caption{ \label{table_bestfits}
   Summary of systematic uncertainties included in the oscillation fit,
   along with the best fit oscillation and systematic parameters 
   returned by each fit.
   For the two-parameter fit, equal oscillation parameters are used 
   for neutrinos and antineutrinos;
   for the four-parameter fit, separate oscillation parameters are used. 
   The best fit systematic parameters are given in units of standard deviations.}
 \end{table*}

\subsection{Results of oscillation fit}
 
\begin{figure}[!tb]
   \includegraphics[width=\columnwidth]{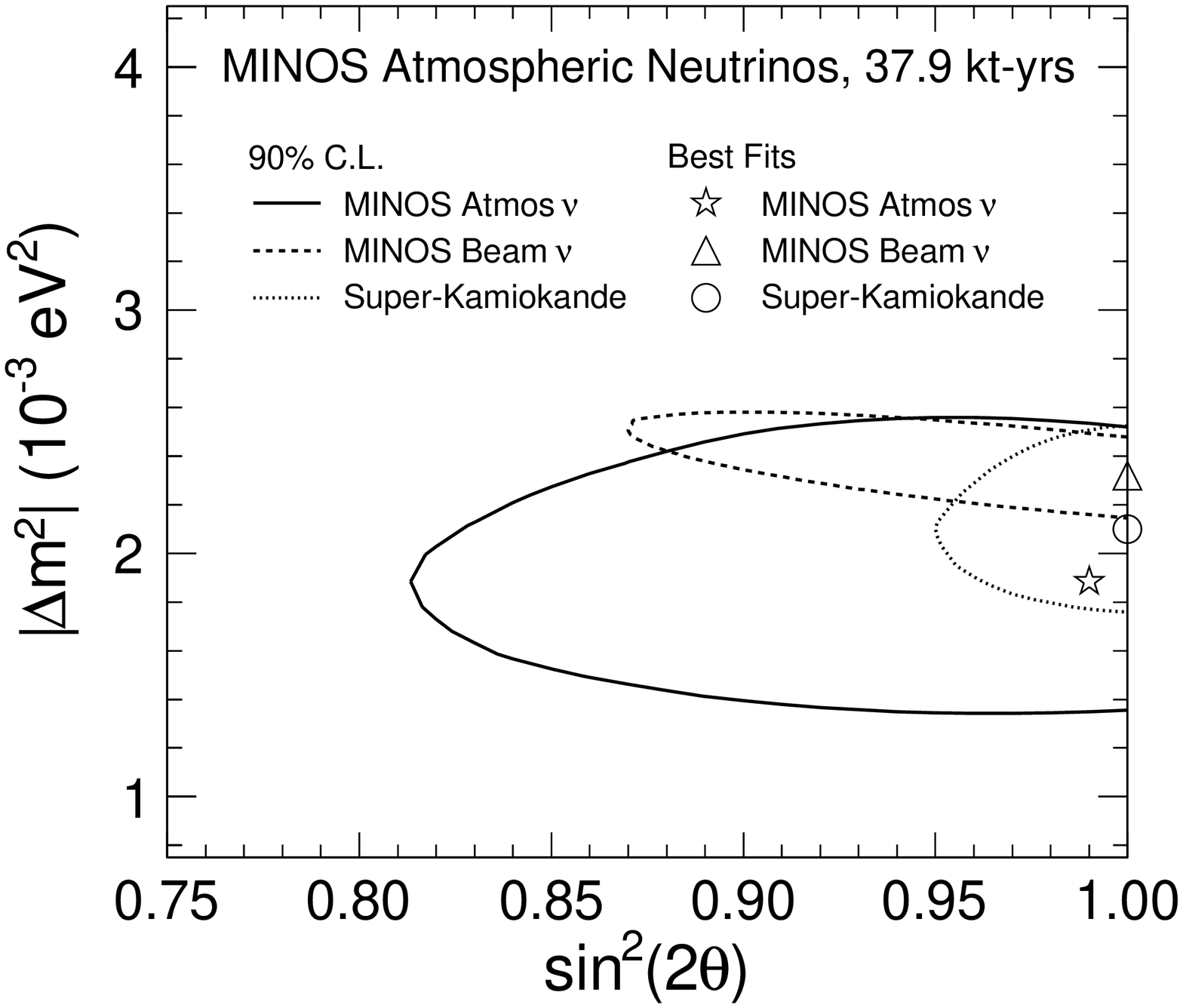}   
   \caption{\label{fig_data_contour_osc_qcomb}
     Confidence limits on the parameters $|\Delta m^{2}|$ and $\mbox{sin}^{2}2\theta$,
     assuming equal oscillations for neutrinos and antineutrinos. The solid line gives
     the 90\% contour obtained from this analysis, with the best fit parameters indicated by the star. 
     For comparison, the dashed line shows the 90\% contour given by the MINOS oscillation 
     analysis of neutrinos from the NuMI beam~\cite{minos3}, with the best fit point indicated 
     by the triangle. The dotted line shows the 90\% contour from the Super-Kamiokande 
     atmospheric neutrino zenith angle analysis (from~\cite{sknubar}), 
     with the best fit point indicated by the circle.}
 \end{figure}

The log-likelihood function is minimized with respect to
the oscillation and nuisance parameters. 
Table~\ref{table_bestfits} summarizes the best fit parameters.
The best fit point occurs at 
($|\Delta m^{2}|$, $\mbox{sin}^{2} 2\theta$)\,=\,($1.9\times 10^{-3}\mbox{\,eV}^{2}$, $0.99$).
The 68\%, 90\% and 99\% confidence limits (C.L.)
on the oscillation parameters are obtained 
in the limit of Gaussian errors
from the locus of points with log-likelihood 
values of $-\Delta \ln \cal{L} =$ (1.15, 2.30, 4.61) 
relative to the best fit point. 
Figure~\ref{fig_data_contour_osc_qcomb} shows the
resulting 90\% contours from this analysis. 
For comparison, this figure also shows the 90\% contours 
from the MINOS beam neutrino analysis~\cite{minos3},
and also from the Super-Kamiokande atmospheric neutrino
zenith angle analysis~\cite{sknubar}.

The log-likelihood surface is used to calculate
single-parameter confidence intervals for each of
the oscillation parameters, by minimizing with 
respect to the other oscillation parameter.
The 90\% single-parameter confidence intervals 
at the best fit point, calculated using this method, are
$|\Delta m^{2}| = (1.9 \pm 0.4 ) \times 10^{-3} \mbox{\,eV}^{2}$
and $\mbox{sin}^{2} 2\theta > 0.86$.
The null oscillation hypothesis is disfavored 
at the level of 9.2 standard deviations.

 \begin{figure*}[!p]
   \includegraphics[width=\textwidth]{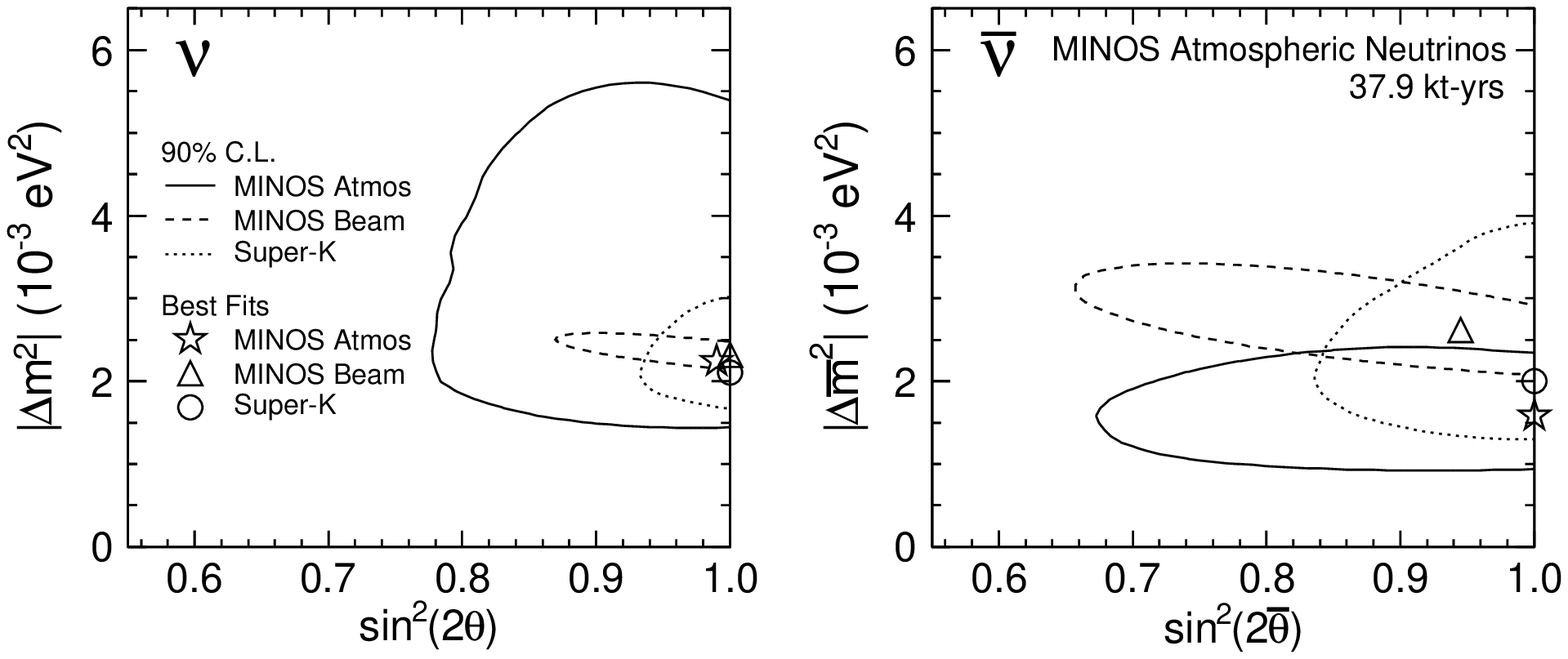}   
   \caption{\label{fig_data_contour_osc_qsep}
     Confidence limits on neutrino (left) and antineutrino (right) oscillation parameters.
     The solid lines show the 90\% contours obtained from this analysis.  
     The two-parameter contours for neutrinos and antineutrinos are calculated by 
     profiling the four-parameter likelihood surface.
     The best fit parameters are indicated by the stars. 
     For comparison, the dashed lines in each plot
     show the 90\% contours from the MINOS analysis of beam data in neutrino mode~\cite{minos3} 
     and antineutrino mode~\cite{minosrhc2}, with the best fit points indicated by the triangles.
     The dotted lines show the 90\% contours from the Super-Kamiokande analysis 
     of atmospheric neutrinos and antineutrinos (from~\cite{sknubar}),
     with the best fit points indicated by the circles.} 
   \vspace{0.25cm}
   \includegraphics[width=\textwidth]{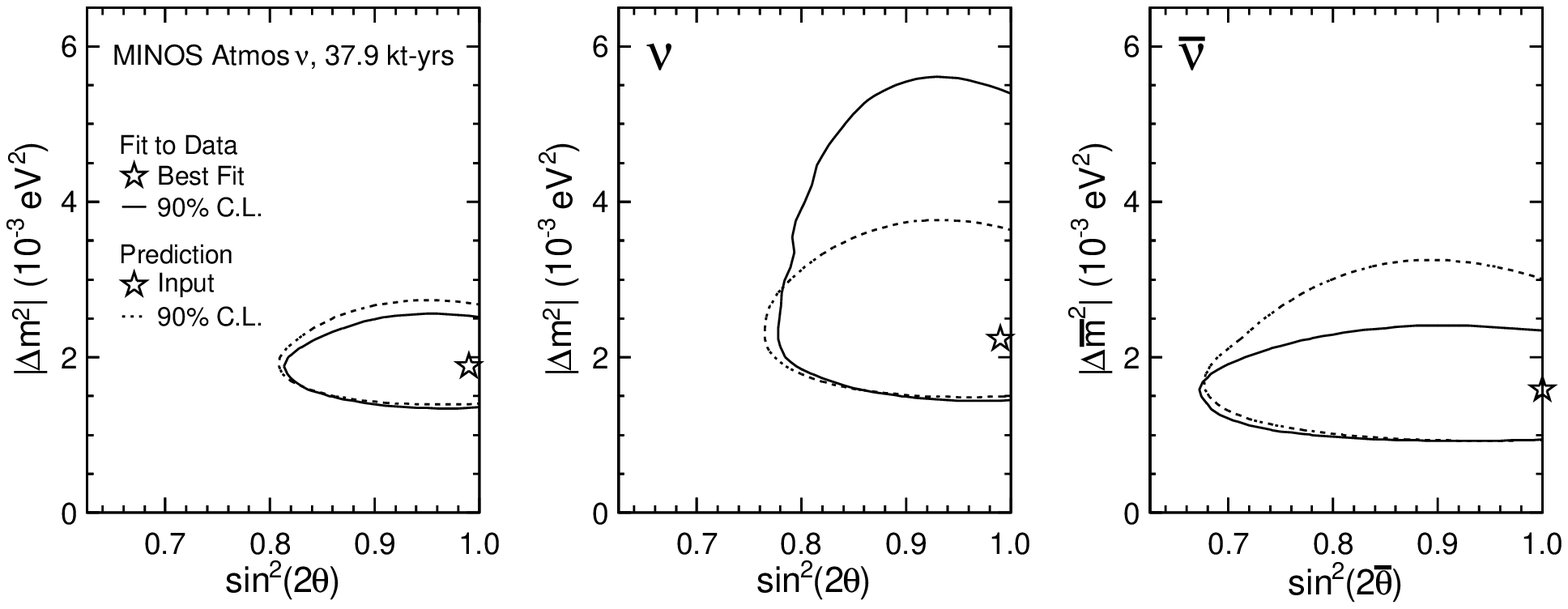}   
   \caption{\label{fig_data_contour_sens}
     Comparisons of the observed and predicted 90\% contours.
     The left plot shows the contours for the two-parameter oscillation fit,
     where neutrinos and antineutrinos take the same oscillation parameters.
     The right two plots, labelled $\nu$ and $\overline{\nu}$, 
     show the contours obtained for neutrinos and antineutrinos, respectively,
     resulting from the four-parameter oscillation fit, where neutrinos and antineutrinos
     take different oscillation parameters. In each case,
     the predicted contours are generated by inputting the best fit parameters
     from the observed data into the simulation, and running the full oscillation fit,
     including the twelve systematic parameters.} 
 \end{figure*} 

 \begin{figure}[!tb]
   \includegraphics[width=\columnwidth]{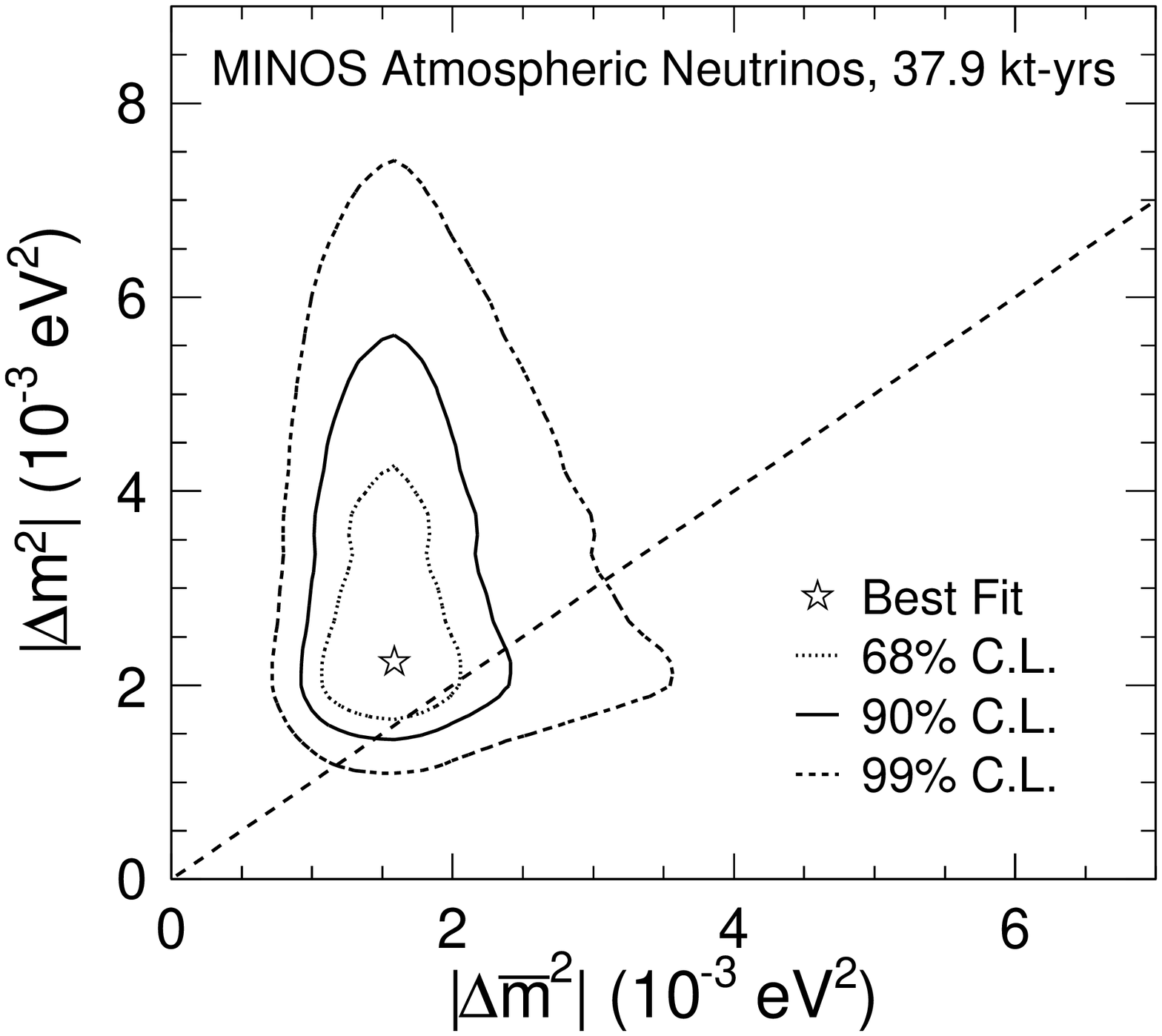}   
   \caption{ \label{fig_data_contour_cpt}
     Confidence limits obtained for the oscillation parameters $|\Delta m^{2}|$ 
     and $|\Delta \overline{m}^{2}|$, representing the mass splittings for neutrinos
     and antineutrinos, respectively. At each point in parameter space, the negative 
     log-likelihood function has been minimized with respect to the mixing parameters
     $\mbox{sin}^{2}2\theta$ and $\mbox{sin}^{2}2\overline{\theta}$.
     The 68\%, 90\% and 99\% contours are indicated by the dotted, solid
     and dashed curves, respectively, and the best fit parameters are indicated by the star. 
     The diagonal dashed line indicates the line of $|\Delta m^{2}|=|\Delta \overline{m}^{2}|$.} 
 \end{figure}

\section{\label{TestingCPT}Fits to neutrino and antineutrino oscillations}

Since the data are separated into pure samples
of neutrinos and antineutrinos, they can be used
to study oscillations separately in neutrinos 
and antineutrinos. The analysis described above 
is extended to incorporate separate oscillation parameters
for neutrinos ($\Delta m^{2},\mbox{sin}^{2} 2\theta$)
and for antineutrinos ($\Delta \overline{m}^{2}$, $\mbox{sin}^{2} 2\overline{\theta}$).
The log-likelihood function is then minimized with 
respect to these oscillation parameters 
and the twelve nuisance parameters. 
The best fit occurs at 
($|\Delta m^{2}|$,\,$\mbox{sin}^{2} 2\theta$)\,=\,($2.2\times 10^{-3}\mbox{\,eV}^{2}$,\,$0.99$)
and ($|\Delta \overline{m}^{2}|$, $\mbox{sin}^{2} 2\overline{\theta}$)\,=\,($1.6 \times 10^{-3} \mbox{\,eV}^{2}$, $1.00$),
as given in Table~\ref{table_bestfits}.
The neutrino and antineutrino oscillation parameters 
are found to be approximately uncorrelated around the best fit point. 
A set of two-parameter profiles can be calculated from the 
four-parameter likelihood surface by minimizing  
with respect to pairs of oscillation parameters.
Figure~\ref{fig_data_contour_osc_qsep} shows the 
resulting 90\% contours obtained for the ($|\Delta m^{2}|$,\,$\mbox{sin}^{2} 2\theta$)
and ($|\Delta \overline{m}^{2}|$,\,$\mbox{sin}^{2} 2\overline{\theta}$) planes.
These results are compared with the 90\% contours from the MINOS analyses
of NuMI beam data acquired in neutrino~\cite{minos3} and antineutrino~\cite{minosrhc2} mode, 
and also the 90\% contours from the SK analysis of 
atmospheric neutrinos and antineutrinos~\cite{sknubar}.

The four-parameter likelihood surface is used to
calculate single-parameter confidence intervals
on each of the four oscillation parameters.
The resulting 90\% C.L. are:
$|\Delta m^{2}| = 2.2^{+2.4}_{-0.6}\times 10^{-3}\mbox{\,eV}^{2}$
and $\mbox{sin}^{2} 2\theta > 0.83$ for neutrinos;
and $|\Delta \overline{m}^{2}| = 1.6^{+0.5}_{-0.5}\times 10^{-3}\mbox{\,eV}^{2}$
and $\mbox{sin}^{2} \overline{\theta} > 0.76$ for antineutrinos.
The null oscillation hypothesis is disfavored
at the level of 7.8 standard deviations for
neutrinos and 5.4 standard deviations
for antineutrinos.

As a measure of the quality of the fit, a set of 10,000 simulated
experiments were generated at the best fit oscillation parameters. 
For each simulated experiment, input systematic parameters were chosen 
from Gaussian PDFs with widths set to the systematic uncertainties.
The best fit parameters were then found for each experiment
by minimizing the log-likelihood function.
For each experiment, the minimum value of $-\Delta \ln \cal{L}$
was recorded; in 22\% of experiments, the value exceeded
that obtained from the fit to the data.

Figure~\ref{fig_data_contour_sens} compares the observed
90\% C.L. from each fit with the predictions from the Monte Carlo simulation, 
calculated by inputting the best fit oscillation parameters into the simulation.
For the two-parameter oscillation fit, where neutrinos and antineutrinos
take the same oscillation parameters, 
there is good agreement between the observed and predicted contours.
For the four-parameter oscillation fit, where neutrinos and antineutrinos
take separate oscillation parameters,
there is a good match between contours
for the limits on the $\mbox{sin}^{2}2\theta$ and $\mbox{sin}^{2}2\overline{\theta}$ parameters 
and the lower limits on the $|\Delta m^{2}|$ and $|\Delta \overline{m}^{2}|$ parameters.
However, the upper limits on these parameters are found to be higher
than predicted for neutrinos and lower than predicted for antineutrinos.

As a check on the observed confidence limits,
the full likelihood surface was calculated
for a set of 250 simulated experiments,
generated at the best fit oscillation parameters
from the two-parameter fit.
The resulting 90\% confidence intervals were then calculated for each experiment. 
In 25\% of these experiments, the confidence intervals obtained for the 
$\Delta m^{2}$ parameter are broader for neutrinos than antineutrinos,
as is the case for the observed data;
in 10\% of the experiments, the relative size of these intervals
is larger than for the observed data. 
These results indicate that the confidence intervals calculated 
from the observed data are reasonable.

Finally, a log-likelihood profile is calculated in the
($|\Delta m^{2}|$, $|\Delta \overline{m}^{2}|$) plane,
by minimizing the log-likelihood function with respect
to the $\mbox{sin}^{2} 2\theta$ and $\mbox{sin}^{2} 2\overline{\theta}$ parameters.
Figure~\ref{fig_data_contour_cpt} 
shows the resulting 68\%, 90\% and 99\% confidence intervals.
This log-likelihood profile is used to place limits 
on the difference between the neutrino and antineutrino
mass splittings $|\Delta m^{2}|$ and $|\Delta \overline{m}^{2}|$.
The single-parameter 90\% confidence intervals, 
assuming Gaussian errors,
are $|\Delta m^{2}|-|\Delta \overline{m}^{2}| = 0.6^{+2.4}_{-0.8} \times 10^{-3} \mbox{\,eV}^{2}$.
This result is consistent with equal mass 
splittings for neutrinos and antineutrinos.

\section{\label{Summary}Summary}

The 5.4~kton MINOS Far Detector has been collecting 
atmospheric neutrino data since August 2003. An analysis 
of the 2553~live-days of data collected up to March 2011 
yields a total of 2072 candidate atmospheric neutrino events. 
The events are separated into 905~contained-vertex muons 
and 466~neutrino-induced rock-muons, produced by 
$\nu_{\mu}$ and $\overline{\nu}_{\mu}$ CC interactions, 
and 701~contained-vertex showers, composed primarily 
of $\nu_{e}$ and $\overline{\nu}_{e}$ CC interactions
and NC interactions.
The curvature of muons in the magnetic field
is used to divide the selected contained-vertex muons and 
neutrino-induced rock-muons into separate samples of 
$\nu_{\mu}$ and $\overline{\nu}_{\mu}$ events. 
The double $\overline{\nu}_{\mu}$/$\nu_{\mu}$ ratio is calculated to be:
 $R^{data}_{\overline{\nu}/\nu}/R^{MC}_{\overline{\nu}/\nu} 
  = 1.03 \pm 0.08\,(\mbox{stat.}) \pm 0.08\,(\mbox{syst.})$.

A maximum likelihood fit to the observed $L/E$ distributions 
is used to determine the atmospheric neutrino oscillation parameters.
The sensitivity to oscillations is improved 
by separating the contained-vertex muons
into bins of $L/E$ resolution, 
and the neutrino-induced rock-muons 
into bins of muon momentum. 
The fit returns 90\% confidence limits of 
$|\Delta m^{2}| = (1.9 \pm 0.4 ) \times 10^{-3} \mbox{\,eV}^{2}$
and $\mbox{sin}^{2} 2\theta > 0.86$.
The oscillation fit is extended to
allow separate oscillation parameters
for neutrinos and antineutrinos.
This fit returns 90\% confidence limits of 
$|\Delta m^{2}|-|\Delta \overline{m}^{2}| = 0.6^{+2.4}_{-0.8} \times 10^{-3} \mbox{\,eV}^{2}$
on the difference between the squared-mass splittings 
for neutrinos and antineutrinos,
consistent with equal mass splittings
for neutrinos and antineutrinos.

\begin{acknowledgments}
This work was supported by the US DOE, the UK STFC, the US NSF, 
the State and University of Minnesota, the University of Athens in Greece, 
and Brazil's FAPESP and CNPq. 
We are grateful to the Minnesota Department of Natural Resources,
the crew of Soudan Underground Laboratory, and the staff of Fermilab,
for their contributions to this effort.
\end{acknowledgments}

\bibliography{atmos_paper_prd}

\end{document}